\documentclass[
preprint,
amsmath,amssymb,
aps,
]{revtex4-1}

\usepackage{graphicx}
\usepackage{dcolumn}
\usepackage{bm}
\usepackage{subfigure}

\usepackage{subfig}

\begin{document}
		

		\title{Gravity-Higgs portal for dark matter decay}

		\author{X. Sun
}
	\email{bestsunxudong@126.com}
\affiliation{Faculty of Science, Kunming University of Science and Technology, Kunming, 650500, China\\
	Yunnan University, Kunming, 650091, China}

		\date{\today}

\begin{abstract}
Dark matter can decay via \emph{gravity portals} and contribute a \emph{minor decay} (MD) width. This paper proposes that nonminimal coupling between the Higgs field and gravity can lead to additional \emph{gravity-Higgs portals} for gravitational dark matter to decay and contribute \emph{additional decay} (AD) width.
For scalar dark matter candidates with a mass less than 5 TeV, the decay branching ratios are almost independent of whether dark matter decay through gravity portals or gravity-Higgs portals; therefore, it is almost impossible to distinguish gravity portals from gravity-Higgs portals by observing decay products.
For fermionic singlet dark matter candidates, their MD products are diverse, while their AD products are monotonous, only neutrinos and Higgs bosons.
As for both dark matter candidates, gravity-Higgs portals prefer to dominate the decay at around several hundred GeV.
Besides, for both TeV dark matter candidates, there is still a vast parameter space alive.
\end{abstract}

		
		
		\maketitle

\section{Introduction}
Dark matter plays a central role in the $\Lambda$-Cold Dark Matter cosmological model~\cite{YangBL,PDM}. Specifically, dark matter is essential for explaining the rotation curves of galaxies, the X-ray behavior of gases from galaxy clusters, the gravitational lensing effect of galaxies and galaxy clusters, the galaxies' redshift surveys, and the anisotropy of cosmic microwave background radiation~\cite{planck2015,planck2018}. Unfortunately, there is no dark matter particle candidate in the standard model (SM) of particle physics~\cite{PDG2016,PDG2018}. These facts imply the existence of physics beyond the SM.

The weakly interacting massive particle is one of the dark matter particle candidates~\cite{VBDM:7,VBDM:10}. If the observed dark matter relic abundance is achieved through the freeze-out mechanism, dark matter should have weak scale interaction strength with the visible sector. Therefore, this dark matter candidate is expected to leave visual signals in cosmic rays~\cite{cholis2021utilizing}, to be detected in colliders or underground scattering experiments~\cite{VBDM:11}. However, no significant signals found so far.
The feebly interacting massive particles is another dark matter particle candidates~\cite{VBDM:12,VBDM:13,VBDM:14,VBDM:15,VBDM:16,VBDM:17}. In this scenario, dark matter particles are produced through the freeze-in mechanism. Unfortunately, the feeble interaction would make it more difficult to find signals from dark matter in detections or experiments.

Dark matter can associate with the visable sector through a variety of portals. For example, the Higgs portals~\cite{ScalarPhantoms,singletscalars,CFportal:4,CFportal:5,CFportal:6,CFportal:7,CFportal:8,CFportal:9,CFportal:10,CFportal:11,CFportal:12,CFportal:13,CFportal:14}, the $Z$ portals~\cite{CFportal:15,CFportal:16,CFportal:17,CFportal:18}, the fermion portals~\cite{Fermionportal:1,Fermionportal:2,Fermionportal:3,Fermionportal:4}, the photon or dark photon portals~\cite{spin1portal:2,spin1portal:3,spin1portal:4} and others~\cite{spin1portal:1,CFportal:21,CFportal:22,spin0portal:1,spin0portal:2,otherportal:1,otherportal:2,otherportal:3}. 
Given that most evidence for the existence of DM comes from its gravitational effects, the unique gravity portal dark matter is widely studied~\cite{Oscar,DMdecayTGP,Sharpfeature,ProbingGDM,VBDM:34,VBDM:35,VBDM:36,VBDM:39,VBDM:41,VBDM:42,VBDM:43,VBDM:44,VBDM:45,VBDM:46,VBDM:47,VBDM:49}. Among the gravity portal dark matter scenarios, dark matter and gravity can be minimally coupled~\cite{VBDM:39,VBDM:42,VBDM:46,VBDM:53,VBDM:54,VBDM:55}, and non-minimally coupled~\cite{barman2021nonminimally,VBDM:59,Oscar,ProbingGDM,SunandDai2}.

	A remarkable feature of dark matter particles is their lifetime. Since they were produced at the time of big bang nucleosynthesis and have survived until the present, their lifetime should be longer than the universe's age. Additionally, photons, neutrinos, or anti-matter produced by dark matter decay cannot exceed the amounts observed by satellite observations. Along this line, the dark matter's lifetime is currently constrained to many orders of magnitude larger than the universe's age~\citep{ReviewIndirectSearch}.

Dark matter can be long-lived for many reasons. Axion's stability can be due to its low mass and weak coupling with the visible sector~\cite{ref:37,ref:38,ref:88,ref:89}. 
	The stability of feebly interacting massive particles and weakly interacting massive particles can be guaranteed by global symmetry~\cite{stabilityofDM,SsymmetryandDM}, under which dark matter particles are charged while all SM particles are neutral.
There can be various discrete symmetries~\cite{DSymmetry:1,DSymmetry:2}, for example, the $\mathbb{Z}_2$ symmetry~\cite{DSymmetry:3,DSymmetry:4,DSymmetry:5}, the $B-L$ or $B+L$ symmetry~\cite{BLSymmetry:1,BLSymmetry:2}. An $\mathbb{Z}_2$ symmetry can be generated dynamically~\cite{PhysRevD.81.015002,PhysRevD.80.085020,PhysRevD.81.075002} or can be accidental~\cite{PhysRevD.82.111701,Cirelli:2009uv,Cirelli:2005uq,Hambye:2008bq}. R-parity is one of $\mathbb{Z}_2$ symmetry. In the MSSM it both stabilizes dark matter and proton~\cite{Rparity:1,Rparity:2,Rparity:3}. There also can be gauge symmetries which stabilize the dark matter~\cite{gaugeSymmetry:2,ref:42,ref:41}.
However, the above stabilizing symmetry is only assumed to exist in the Minkowski spacetime. In reality, both the dark sector and the visible sector are affected by gravity. Moreover, gravitational effects can violate global symmetries~\citep{GandGSym,ref:44}. In some cases, $\mathbb{Z}_2$ symmetry broken operator can exist~\cite{ref:45,ref:46,ref:47,ref:48}, implying that dark matter particles can be unstable in curved spacetime.

	\citet{Oscar} first proposed that gravity can violate the global symmetry through operators linear to the dark matter field that is coupled nonminimally to gravity through the Ricci scalar (i.e., gravity portals); such coupling can, in turn, lead to their decay. This decay phenomenology has been subsequently studied in detail. For example, decay branching ratios of several widely studied dark matter candidates were given by \citet{DMdecayTGP}. In addition, the photon spectra produced by the decay of scalar singlet dark matter with masses less than 1~GeV would show a sharp feature~\citep{Sharpfeature}.
Still, there is another gravity portal scenario—for example, the affine gravitational scenario for dark matter decay~\cite{PhysRevD.102.084036}. Specifically, dark matter can couple to the spacetime affine connection through a $\mathbb{Z}_2$-symmetry breaking term, resulting in dark matter decay.

	One of the most critical features of dark matter decay through gravity portals is that it will decay into any particle subjected to gravitational interactions as long as it is kinematically allowed. Therefore, the coupling way and the coupling strength between SM particles and gravity are important factors determining the decay phenomenology of gravity portal dark matter. Thus, coupling between SM particles and gravity also needs careful consideration, and such couplings may not necessarily be minimal. Based on the following two reasons, this paper considers the effect of the coupling between the Higgs field and the Ricci scalar on the decay of gravitational dark matter. One reason is that the coupling between a scalar field and the Ricci scalar is expected to exist for renormalizing scalar interactions in curved space~\cite{Parker09,renormalizingcurved1,renormalizingcurved2,renormalizingcurved3,renormalizingcurved4,renormalizingcurved5,renormalizingcurved6,renormalizingcurved7}. Another reason is that of all the allowed nonminimal couplings between SM particles and gravity, the only dimension-four operator is the coupling between the Higgs field and the Ricci scalar~\citep{BoundsHiggsGravity}, which has particle physical and cosmological implications. For example, it can slightly change the weak gauge boson scattering process, which can be tested via the Large Hadron Collider (LHC)~\cite{JFandEF,GravityHiggsWeakBoson}. The coupling between the Higgs field and the Ricci scalar is also essential for gravitational dark matter to produce the observed dark matter thermal relic abundance, and which also allows scalar gravitational dark matter to become feasibly testable~\citep{ProbingGDM}. A reasonable conjecture is that such nonminimal coupling could enlarge the decay width of dark matter, and in some cases, the AD rate would dominate the decay phenomenology of dark matter. Nonminimal coupling operators of other SM fields to gravity are dimension-six or above. Therefore, their influences will be suppressed by a unification mass scale, so this work neglect them.

	The remainder of this paper is structured as follows. Section~\ref{sec:reviewGPDM} gives a brief review on dark matter decaying via gravity portals. Section~\ref{DMandHiggscoupletoR} briefly introduces the gravity-Higgs portals for dark matter to decay. Sections~\ref{GravityInducedMixing} and \ref{FermionicDMdecay} investigate in detail the decay phenomenology of scalar singlet dark matter and fermionic singlet dark matter, respectively, including their branch ratios, decay widths, and set constraints on nonminimal coupling constants. Finally, in Section~\ref{TheSummary} a summary of this work is provided.

\section{Review of dark matter decay through gravity portals}\label{sec:reviewGPDM}
This section briefly reviews the progress that dark matter decays from nonminimal coupling to gravity. It is firstly proposed by \citeauthor{Oscar} that gravity can break the global symmetry, which guarantees the stability of dark matter and then lead to the decay of dark matter~\citep{Oscar}. Specifically, the global symmetry can be broken by an operator linear to the Ricci scalar:
\begin{equation}
	-\xi_\varphi R F(\varphi,X)
\end{equation}
with $\xi_\varphi$ the coupling constant, $R$ the Ricci scalar, $F(\varphi,X)$ the function of the dark matter field $\varphi$ and SM fields $X$.
This operator induces the decay of dark matter to gravitons with decay rate scale with $(\xi_\varphi\kappa^2)^2$ where $\kappa=\sqrt{8\pi G}$ is the inverse (reduced) Planck mass and $G$ is the Newtonian gravitational constant. This coupling also induces the decay of dark matter to all the SM particles, with gravitons acting as a mediator since all the SM particles are subject to gravitational interactions. In this scenario, all the SM particles couple to gravity minimally. This paper calls this part of the decay rate of dark matter the minor decay (MD) rate. Figuratively, this paper subscript ``MD'' to the effective interactions shown by Eq.~\ref{eq:minimalint}. \citeauthor{Oscar} think that the resulting effective interactions can be obtained through a Weyl transformation from the Jordan Frame to the Einstein Frame~\citep{Oscar,DMdecayTGP},
\begin{equation}
	\tilde{g}_{\mu\nu} = \Omega^2 g_{\mu\nu},~\Omega^2=1+2\xi_\varphi\kappa^2 F(\varphi,X).
\end{equation}
\begin{figure}[!htb]
	\subfigure[]{\includegraphics[width=0.45\textwidth]{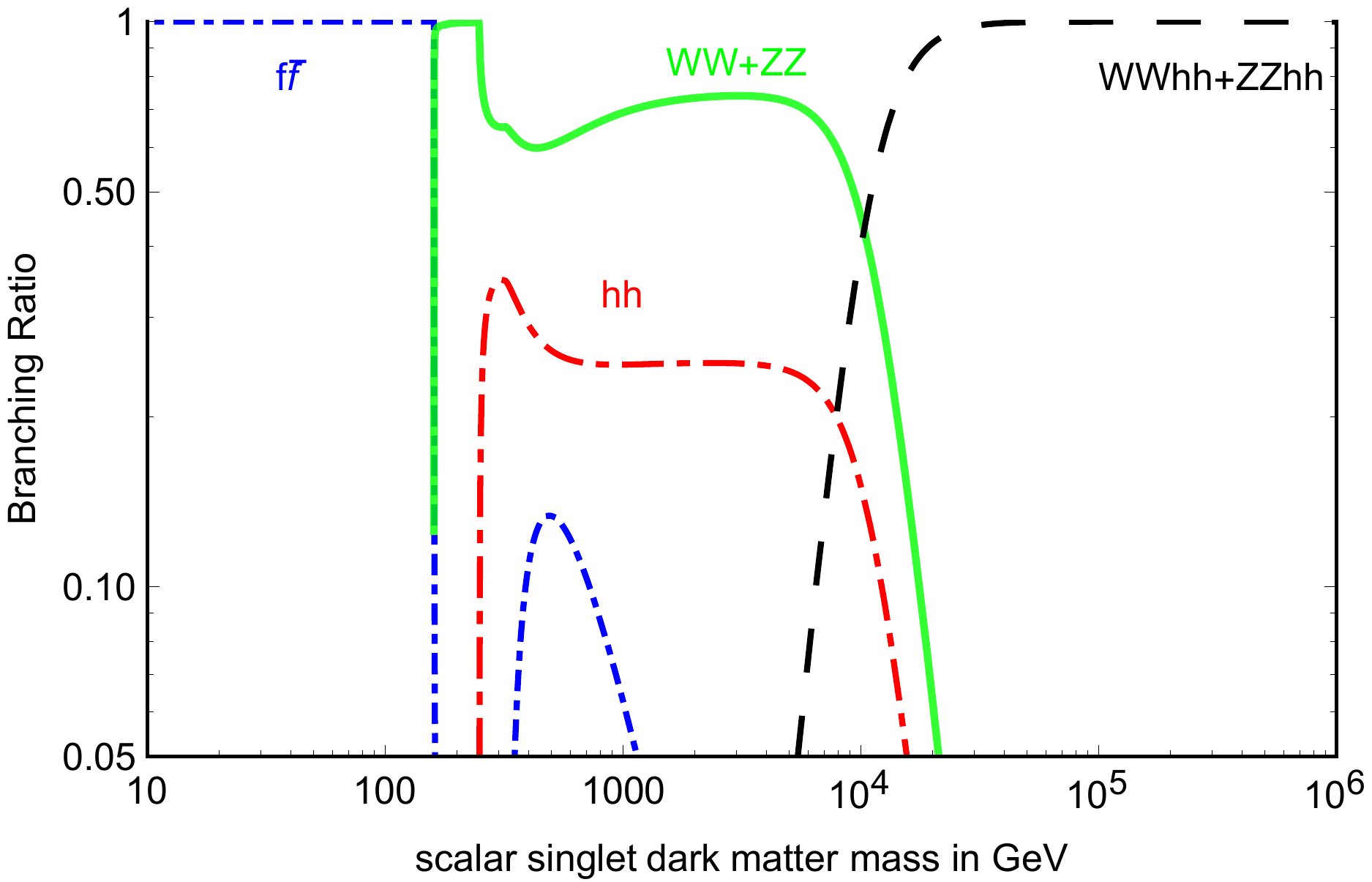}} 
	\subfigure[]{\includegraphics[width=0.45\textwidth]{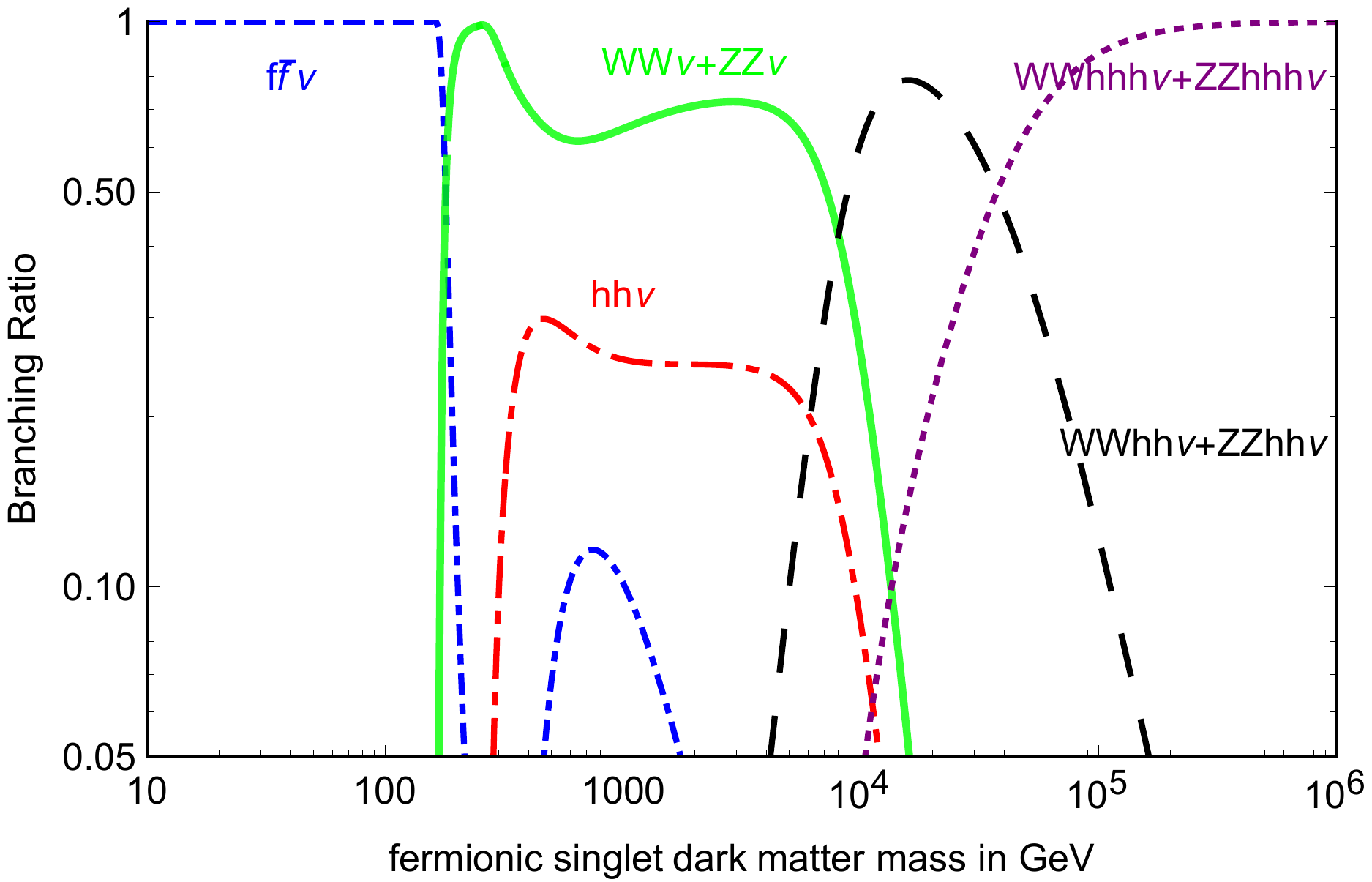}} 
	\caption{\label{fig:correctedBR}
		(a) Decay branch ratios of scalar singlet dark matter candidate $\phi$ via the global $\mathbb{Z}_2$ symmetry breaking operator $-\xi_\varphi \phi R$. In the figure, f, h, W, Z denotes the fermions, the Higgs bosons, the W bosons, and the Z bosons, respectively.\\
		(b) Decay branch ratios of fermionic singlet dark matter candidate $\chi$ via the global $\mathbb{Z}_2$ symmetry breaking operator $-\xi_\varphi (\bar{l}\breve{H}\chi+\bar{\chi}\breve{H}^\dag l)R$, where $\breve{H}=i\sigma_2 H^*$, $l$ stands for electroweak lepton doublet, $\boldsymbol{\sigma}=(\sigma_1,\sigma_2,\sigma_3)$ are the Pauli matrices, $\nu$ denotes the neutrinos.
	}
\end{figure}
However,~\citeauthor{bezrukov2020miracle} point out that the fermionic fields should also be conformally rescaled~\cite{bezrukov2020miracle},
\begin{equation}
f^{\text{(EF)}}=\Omega^{-\frac{3}{2}}f^{\text{(JF)}},
\label{eq:rescalefermions}
\end{equation}
where $f$ denotes the fermionic fields, the superscript EF denotes the Einstein Frame, the superscript JF denotes the Jordan Frame. In the rest of this paper, I drop all the superscript of the fermionic fields.
After Weyl transformation and conformal rescaling of the fermionic fields, the resulting effective interactions should be,
\begin{equation}
	\mathcal{L}_{\text{MD}}^{\text{(EF)}}=
	\Omega^{-2} \tilde{\mathcal{T}}_H+\Omega^{-1}\mathcal{L}_Y-\Omega^{-4}\mathcal{V}_H,
	\label{eq:minimalint}
\end{equation}
where $\tilde{\mathcal{T}}_H$ is the kinetic term of the Higgs field, $\mathcal{L}_Y$ are the Yukawa terms and $\mathcal{V}_H$ is the Higgs potential.

Expanding Eq.~\ref{eq:minimalint} to the first order of $\varphi$ we get
\begin{equation}
\mathcal{L}_{\text{MD},\varphi}^{\text{(EF)}}=
-\xi_\varphi\kappa^2 \frac{\partial F}{\partial \varphi}\bigg|_{\varphi=0} \varphi
[
2\tilde{\mathcal{T}}_H+ \mathcal{L}_Y-4\mathcal{V}_H
]
.
\label{eq:treeleveldecay}
\end{equation}
In the above equation, the coupling strength between dark matter and the SM particles is proportional to $\xi_\varphi\kappa^2$.
Following the procedure and partial results outlined by~\citet{DMdecayTGP}, Fig.~\ref{fig:correctedBR} (a) and Fig.~\ref{fig:correctedBR} (b) give the corrected decay branch ratios of the scalar singlet dark matter candidate $\phi$ and the fermionic singlet dark matter candidate $\chi$ respectively.
In the figure, the scalar field $H$ is parameterized as
$
H=
\frac{1}{\sqrt{2}}
\begin{pmatrix}
0 \\
v+h
\end{pmatrix}
,
$
$v=246$ GeV is the Higgs field's vacuum expectation value, and $h$ is the Standard Model Higgs boson.

\section{Gravity-Higgs portals for dark matter to decay}\label{DMandHiggscoupletoR}

\subsection{The framework}

This section proposes additional gravity-Higgs portals for dark matter to decay.
A straightforward way for dark matter and the Higgs field to couple to gravity nonminimally is through coupling to the Ricci scalar:
	\begin{equation}
	\mathcal{S}^{(\text{JF})}
	=
	\int d^4x \sqrt{-g}
	[
	-\frac{R}{2\kappa^2}
	-\xi_\varphi F(\varphi,X) R
	- \xi_h H^\dag H R
	]
	\label{eq:FRandHR}
	\end{equation}
where $g$ is the determinant of the metric tensor $g_{\mu\nu}$, and $R$ is the Ricci scalar. $\xi_\varphi F(\varphi, X) R$ is the nonminimal coupling term that can break the global symmetry. In other words, dark matter can decay into gravitons through it. $F(\varphi, X)$ is a real function of the dark matter field $\varphi$, and possibly other fields $X$, and $\xi_\varphi$ is a coupling constant. $\xi_h H^\dag H R$ describes the nonminimal coupling between the Higgs field and gravity, where $H$ is the Higgs doublet, and $\xi_h$ is the coupling constant.
Together with $\xi_\varphi F(\varphi, X) R$, $\xi_h H^\dag H R$ allow the dark matter to decay into the Higgs bosons with gravitons acting as a mediator. Section~\ref{GravityInducedMixing} will show that these two operators also allow the dark matter to decay into other particles coupling to the Higgs bosons with gravitons and Higgs bosons acting as mediators when the dark matter candidate is a scalar singlet.
Given that this part of the decay rate of dark matter originates from the nonminimal coupling between the Higgs field and gravity, This paper calls this part of the decay rate the additional decay (AD) rate. Figuratively, This paper subscript ``AD'' to the effective interactions shown by Eq.~\ref{eq:enhancedterm}.

The Higgs field gains a vacuum expectation value after electroweak symmetry breaking, $v=246~\text{GeV}$, which results in the following relation between $\kappa$ and $\xi_h$ that $\kappa^{-2}+\xi_h v^2= M_P^2/8\pi$,
where $M_P=1.22\times10^{19}~\text{GeV}$ is the Planck mass.

	Nonminimal coupling between the Higgs field and gravity will affect the production and decay rates of the Higgs boson, which can be tested at the LHC. Along this line, the strength of nonminimal coupling between the Higgs field and gravity has been constrained to~\cite{BoundsHiggsGravity}
	\begin{equation}
		|\xi_h|<2.6\times 10^{15}
		.
	\end{equation}
	 Naturally, $\xi_h\ll M_P^2/(8\pi v^2)\simeq10^{32}$ so that the approximation $\kappa=\sqrt{8\pi}M_P^{-1}$ can be used.
	
	By performing the Weyl transformation of 
		\begin{equation}
	\tilde{g}_{\mu\nu}=\Omega^2 g_{\mu\nu}
	\label{WeylTrans}
	\end{equation}
on the metric tensor, the dark matter fields and Higgs fields can be decoupled from the Ricci scalar in the Einstein Frame as~\cite{Oscar,JFandEF,SunandDai}:
	\begin{equation}
\mathcal{S}^{(\text{EF})}
=
\int d^4x \sqrt{-\tilde{g}} [
-\frac{\tilde{R}}{2\kappa^2}
+\frac{3}{\kappa^2} \frac{\Omega_{,\rho}\tilde{\Omega}^{,\rho}}{\Omega^2}
]
\label{actioninEF}
\end{equation}
where $\Omega^2=1+2\kappa^2 \xi_h H^\dag H+2\kappa^2 \xi_\varphi F(\varphi,X)$, and all quantities with a tilde are formed by $\tilde{g}^{\mu\nu}$.
	
	The second term on the right side of Eq.~\ref{actioninEF} contains the effective interactions between dark matter and the Higgs field,
	which can be expanded as:
	\begin{eqnarray}
	\mathcal{L}^{\text{(EF)}}_{\text{AD}}
	=
	\frac{3}{\kappa^2} \frac{\Omega_{,\rho}\tilde{\Omega}^{,\rho}}{\Omega^2}
	=
	\frac{3 \kappa^2}{\Omega^4}
	[\xi_h\xi_\varphi \tilde{F}^{,\rho}(\varphi,X) (H^\dag H)_{,\rho}
	+\xi_h^2 \tilde{g}^{\mu\nu} (H^\dag H)_{,\mu}  (H^\dag H)_{,\nu}
	+\xi_\varphi^2 \tilde{F}^{,\rho}(\varphi,X) F_{,\rho}(\varphi,X)].   \nonumber\\
	\label{eq:enhancedterm}
	\end{eqnarray}
The first term on the right-hand side reveals that the coupling strength between dark matter and the Higgs field is proportional to $\xi_\varphi\xi_h\kappa^2$ in the gravity-Higgs scenario. In comparison, Eq.~\ref{eq:treeleveldecay} shows that the coupling strength between dark matter and the SM particles is proportional to $\xi_\varphi\kappa^2$ in the gravity portal scenario. So we can conclude that the value of $\xi_h$ determines which channel (AD or MD or both channels) dominates dark matter's decay phenomenology.

\subsection{Unitarity bounds on the coupling of the Higgs field to the Ricci scalar}

		This section derives the unitarity bounds on $\xi_h$. In the unitary gauge, the second term to the right side of Eq.~\ref{eq:enhancedterm} can be expressed as 
	$
		3\kappa^2
		(
		2 \xi_h^2 v h h_{,\rho}\tilde{h}^{,\rho}
		+ \xi_h^2 h^2 h_{,\rho}\tilde{h}^{,\rho}
		+ \xi_h^2 v^2 h_{,\rho}\tilde{h}^{,\rho}
		)/\Omega^4
	$. It reveals that the elastic scattering process $h, h\to h ,h$ could exist. The corresponding vertex rule is $12i\xi_h^2\kappa^2(-p_{\text{in1}}p_{\text{in2}}-p_{\text{out1}}p_{\text{out2}}+p_{\text{in1}}p_{\text{out1}}+p_{\text{in1}}p_{\text{out2}}+p_{\text{in2}}p_{\text{out1}}+p_{\text{in2}}p_{\text{out2}})$, where $p$ denotes a four momenta of a Higgs boson, the subscript ``in'' and ``out'' denotes the inital state of a Higgs boson and the final state of a Higgs boson respectively.

		For the $h,h\to h,h$ process, the invariant amplitude is
	\begin{subequations}
		\begin{equation}
		\mathcal{M}=24\xi_h^2\kappa^2m_\phi^2
		.
		\end{equation}
		Then I can impose the partial wave unitarity condition, $\mathsf{Re}\{a_0\}<1/2$, where
		\begin{equation}
		\mathcal{M}(\theta)=16\pi\sum_{j=0}^{\infty}a_j(2j+1)P_j(\text{cos }\theta)
		\end{equation}
		with $P_j(\text{cos }\theta)$ being the Legendre polynomials that satisfy $P_j(1)=1$.
		We get
		\begin{equation}
		|\xi_h|<
		\sqrt{\frac{\pi}{3}}
		\frac{1}{\kappa m_h}
		.
		\end{equation}
	\end{subequations}
		Substituting $m_h=125~\text{GeV}$ into the equation above, we get the unitarity bounds on the coupling of the Higgs field to the Ricci scalar:
		\begin{equation}
		|\xi_h|<2\times10^{16}
		.
		\end{equation}

\section{Scalar singlet as dark matter candidate}\label{GravityInducedMixing}

A scalar singlet is the most straightforward candidate for dark matter. It could couple to gravity nonminimally and obtain its relic density through the freeze-out mechanism~\cite{SunandDai2}.
For scalar singlet dark matter (denoted by $\phi$) that is charged under a discrete $\mathbb{Z}_2$ symmetry, under which $\phi$ is odd $(\phi\to-\phi)$ and all SM fields are even~\cite{singletscalars,ScalarPhantoms,updateSSDM}, it could set $F(\varphi, X)=\phi$ so that the hypothetical $\mathbb{Z}_2$ symmetry is broken~\citep{Oscar}.

	Let us focus on the effective interactions $3\kappa^{-2} \Omega_{,\rho}\tilde{\Omega}^{,\rho} \Omega^{-2} $. Working with the unitary gauge, after electroweak symmetry breaking, the Weyl factor becomes $\Omega^2=1+\kappa^2 \xi_h (v+h)^2 +2\kappa^2 \xi_\varphi \phi$, where $h$ is the SM Higgs boson. The effective interactions can be expressed as:
	\begin{eqnarray}
	\frac{3\kappa^2}{\Omega^4}
	(
	2 \xi_h \xi_\varphi v h_{,\rho}\tilde{\phi}^{,\rho}
	+2 \xi_h \xi_\varphi h h_{,\rho}\tilde{\phi}^{,\rho}
	+2 \xi_h^2 v h h_{,\rho}\tilde{h}^{,\rho}
	+ \xi_h^2 h^2 h_{,\rho}\tilde{h}^{,\rho}
	+ \xi_h^2 v^2 h_{,\rho}\tilde{h}^{,\rho}
	+ \xi_\varphi^2 \phi_{,\rho}\tilde{\phi}^{,\rho}
	)
	\label{eq:mixingterms}
	\end{eqnarray}
       The first term is a kinetic mixing term between the scalar dark matter and the Higgs field, which indicates that after the Higgs field gains the vacuum expectation value, $\phi$ and $h$ are no longer eigenstates of mass. This kinetic mixing term corresponds to a physical picture that while gravity and Higgs bosons act as messengers, dark matter decay to any SM particles that couples to the Higgs boson. So, we should find their mass eigenstates to investigate their properties further. The second term is similar to Yukawa coupling, which contributes to the decay channel, $\phi\to h,h$, and cannot be neglected when calculating the decay rate. The following two terms are self-interactions of the Higgs boson, which are suppressed by $\kappa^2$. Compared with self-coupling in the SM, they are neglectable. The last two terms indicate that the scalar dark matter and the Higgs boson should be rescaled in the Einstein Frame. Since they are also suppressed by $\kappa^2$, we could also neglect these effects.

\paragraph{The mass basis}
	
First, let us investigate the physics of the first term of Eq.~\ref{eq:mixingterms}. Let us focus on the following Lagrangian:
	\begin{eqnarray}
	\frac{1}{2} h_{,\mu}\tilde{h}^{,\mu}-\frac{1}{2}m_h^2 h^2
	+
	\frac{1}{2} \phi_{,\mu}\tilde{\phi}^{,\mu}-\frac{1}{2}m_\phi^2 \phi^2
	+
	\alpha h_{,\mu}\tilde{\phi}^{,\mu}
	\end{eqnarray}
where $\alpha=6\xi_\varphi\xi_h\kappa^2v$ is dimensionless, noting that the last term is the leading order term of the kinetic mixing term of Eq.~\ref{eq:mixingterms}.
	
	I find it should make the following rotation and rescaling operations:
	\begin{subequations}
		\begin{equation}
		m_h h= m_{h'} h' \text{cos}\theta - m_{\phi'} \phi' \text{sin} \theta
		\label{RotationandRescaling_A}
		\end{equation}
		\begin{equation}
		m_\phi \phi= m_{h'} h' \text{sin}\theta + m_{\phi'} \phi' \text{cos} \theta
		\end{equation}
		\label{RotationandRescaling}
	\end{subequations}
where:
	\begin{subequations}
		\begin{equation}
		\text{tan}2\theta=
		\frac{2\alpha m_h m_\phi}{m_\phi^2-m_h^2}
		=2\alpha\frac{m_h}{m_\phi}(1-\frac{m_h^2}{m_\phi^2})^{-1}
		\end{equation}
		\begin{equation}
		m_{h'}=
		[\frac{1}{2}(\frac{1}{m_h^2}+\frac{1}{m_\phi^2})
		+\frac{\text{cos}2\theta}{2}(\frac{1}{m_h^2}-\frac{1}{m_\phi^2})
		+\frac{\alpha\text{sin}2\theta}{m_h m_\phi}]^{-1/2}
		\end{equation}
		\begin{equation}
		m_{\phi'}=
		[\frac{1}{2}(\frac{1}{m_h^2}+\frac{1}{m_\phi^2})
		-\frac{\text{cos}2\theta}{2}(\frac{1}{m_h^2}-\frac{1}{m_\phi^2})
		-\frac{\alpha\text{sin}2\theta}{m_h m_\phi}]^{-1/2}
		,
		\end{equation}
	\end{subequations}
	The focused Lagrangian then can be expressed in the mass basis:
	\begin{eqnarray}
	\frac{1}{2}
	h'_{,\mu}\tilde{h'}^{,\mu}
	-
	\frac{1}{2}
	m_{h'}^{2}
	h'^2
	+
	\frac{1}{2}
	\phi'_{,\mu}\tilde{\phi'}^{,\mu}
	-
	\frac{1}{2}
	m_{\phi'}^{2}
	\phi'^2
	\end{eqnarray}
where we should notice that $\phi'$ is the mass eigenstate of the scalar dark matter, and $h'$ is the detected Higgs boson by the LHC.

\subsection{Branch ratios}

	\begin{table}[b]
		\caption{\label{vertexrules}
			Feynman rules for the decay of scalar singlet dark matter.
			\footnote{In the Feynman rules, the symmetry factors are taken into account.}
		}
		\begin{ruledtabular}
			\begin{tabular}{lcdr}
				\textrm{Physical process\footnote{In the table, $f$ denotes fermions, $W$ and $Z$ represents the $W$ boson and the $Z$ boson respectively. Besides, $m_a$ refers to the mass of the particle $a$. All channels are contributed by the first term of Eq.~\ref{eq:mixingterms} (more specifically, Eq.~\ref{RotationandRescaling_A}) except for the last channel, where $\phi'\to h',h'$ is partially contributed by the second term of Eq.~\ref{eq:mixingterms}.}}&
				\textrm{Feynman rules\footnote{In the table, $m_Y$ represents the mass of the $W$ boson or the $Z$ boson.}}\\
				\colrule
				$\phi' \rightarrow \bar{f}  f$
				& $6i\xi_\varphi\xi_h\kappa^2\frac{m_{\phi'}^2 m_f}{m_{\phi'}^2-m_{h'}^2} $ \\
				\hline
				$ \phi' \to W W ;(\phi' \to Z Z) $
				& $
				-12i\xi_\varphi\xi_h\kappa^2\frac{m_{\phi'}^2 m_Y^2}{m_{\phi'}^2-m_{h'}^2 } \tilde{g}_{\mu\nu}
				$ \\
				\hline
				$ \phi' \to W W h' ;(\phi' \to Z Z h') $
				& $
				-6i\xi_\varphi\xi_h\kappa^2\frac{m_{\phi'}^2 m_Y^2}{(m_{\phi'}^2-m_{h'}^2)v } \tilde{g}_{\mu\nu}
				$\\
				\hline
				$ \phi'\to h' h' h'$
				& $
				18i\xi_\varphi\xi_h\kappa^2
				\frac{m_{\phi'}^2 m_{h'}^2}{(m_{\phi'}^2-m_{h'}^2)v }
				$ \\
				\hline
				$ \phi'\to h' h'$
				& $
				6i\xi_\varphi\xi_h\kappa^2
				(
				\frac{3 m_{\phi'}^2 m_{h'}^2}{m_{\phi'}^2-m_{h'}^2 }
				+p_{\phi'}\cdot p_{h'1}+p_{\phi'}\cdot p_{h'2}
				)
				$ \\
			\end{tabular}
		\end{ruledtabular}
	\end{table}

	\begin{table}[t]
		\caption{\label{decaymodes}
			Tree-level decay modes of the scalar singlet.
		}
		\begin{ruledtabular}
			\begin{tabular}{lcdr}
				\textrm{Decay mode}&
				\textrm{Asymptotic scaling}\\
				\colrule
				$\phi'\to h'h', WW, ZZ$
				& $m_{\phi'}^3$ \\
				$\phi'\to f\bar{f}$
				& $m_{\phi'} m_f^2$ \\
				$\phi'\to h'h'h'$
				& $m_{\phi'} m_{h'}^4/ v^2$ \\
				$\phi'\to WWh',ZZh'$
				& $m_{\phi'}^5/ v^2$ \\
			\end{tabular}
		\end{ruledtabular}
	\end{table}

In the $\alpha\to0$ limit, the rotation angle, $\theta$, is infinitesimal, and there are many useful approximations that can be applied: $m_{h'}\simeq m_h$, $m_{\phi'}\simeq m_\phi $, and $\text{sin}\theta \simeq \theta \simeq\alpha m_h m_\phi/(m_\phi^2-m_h^2) $. Eq.~\ref{RotationandRescaling_A} shows that particles that couple to the Higgs boson are also coupled to the scalar singlet $\phi'$. The couplings could be obtained simply through an approximated replacement, $h\to h'-\phi' \theta m_{\phi'}/m_{h'}$. It should also be noticed that the scalar singlet $\phi'$ couples to SM particles through the Higgs field, $h$. Moreover, the rotation angle turns out to be a key parameter for the decay of the scalar singlet.

\begin{figure}[b]
	\includegraphics[scale=0.7]{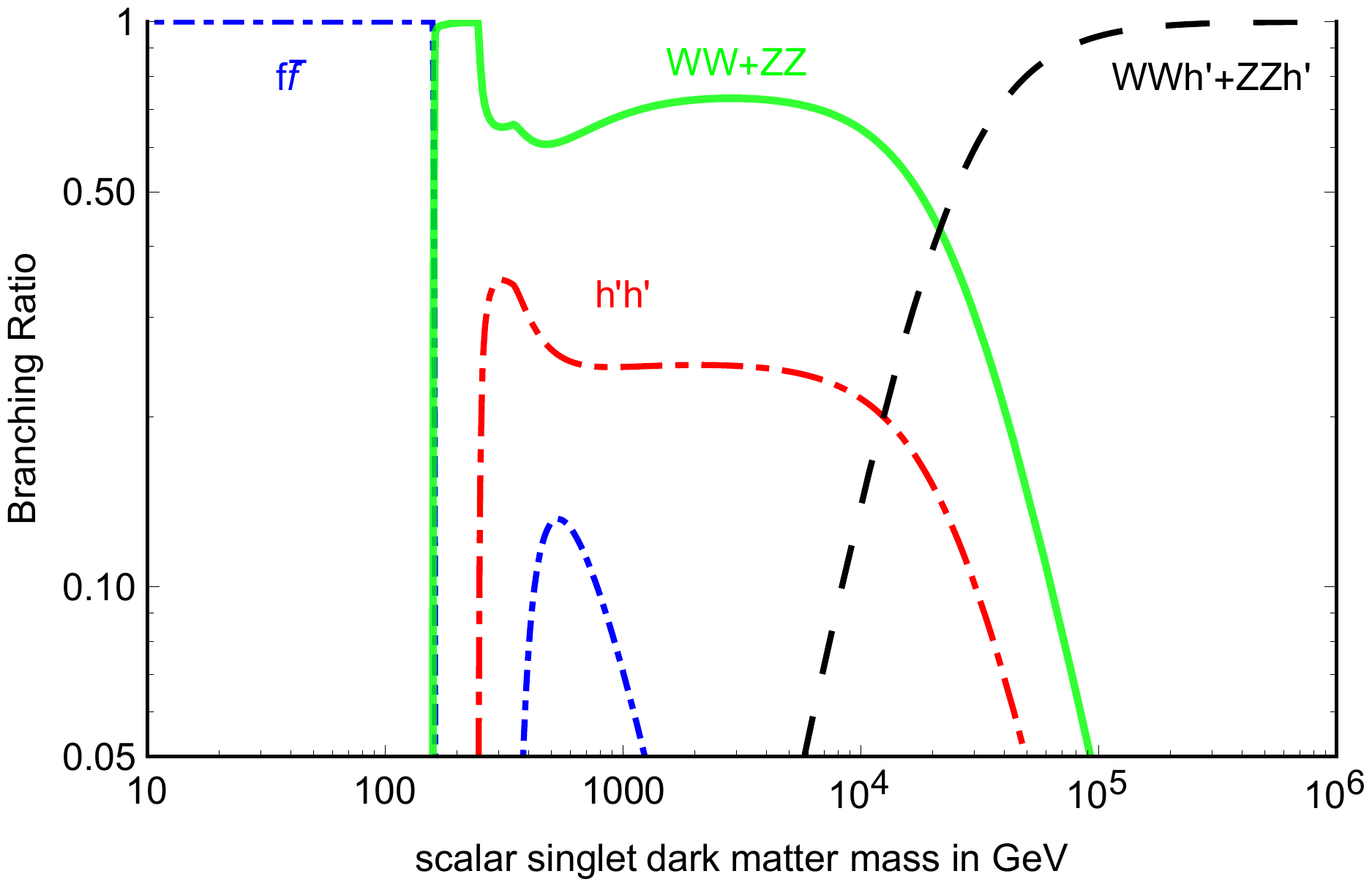}
	\caption{\label{fig:BR} Decay branching ratios for the scalar singlet dark matter as a function of its mass.}
\end{figure}

	The complete set of decay channels with corresponding Feynman rules are given in Table~\ref{vertexrules}. All of the couplings at the tree level arise from kinetic mixing between the scalar singlet dark matter and the Higgs field, except one that arises from the second term of Eq.~\ref{eq:mixingterms}. In Table~\ref{decaymodes}, all possible tree-level decay channels are listed, together with the asymptotic scaling of large dark matter masses. Also, figure~\ref{fig:BR} shows the exact decay branch ratios of the scalar singlet dark matter.

	For scalar singlet masses below 160 GeV, the decay rate is dominated by $\phi'\to f \bar{f}$, since it is the only channel allowed by the phase space. Once the $\phi'\to b\bar{b}$ channel opens (where $b$ denotes the bottom quark), the decay branch ratio of the scalar singlet will be similar to that of the Higgs boson due to kinetic mixing between the scalar singlet and the Higgs field; specifically, it is dominated by the $\phi'\to b\bar{b}$ channel. For intermediate masses, $160~\text{GeV}\leq m_{\phi'}\leq 10^4~\text{GeV}$, decay channels with high thresholds, $\phi'\to WW,ZZ,h'h',t\bar{t}$, where $t$ is the top quark, dominate the decay rate. Compared with the other two-body decay channels, the $\phi'\to t\bar{t}$ channel is suppressed by $m_f^2/m_{\phi'}^2$ so that a hump between 400 GeV and 1000 GeV forms. Meanwhile, three-body decays are suppressed by the smaller phase space. For large scalar singlet masses, compared with the $\phi'\to WW,ZZ,h'h'$ channels, the $\phi'\to WWh',ZZh'$ channels receive a relative enhancement factor $m_{\phi'}^2/v^2$, meaning they dominate the decay rate. Specifically, the dominant mode consists of decays with longitudinal components. Note that the $\phi'\to h'h'h'$ channel is highly suppressed and cannot be displayed in Figure~\ref{fig:BR}.

	The analytical expressions of the decay rates of two-body decay channels are:
	\begin{subequations}
		\begin{equation}
		\Gamma_{\phi'\to f\bar{f}}
		=
		N^{(f)}
		\frac{9\xi_\varphi^2 \xi_h^2 \kappa^4}{2\pi}
		m_{\phi'}^3 x_f
		\frac{(1-4x_f)^{3/2}}{(1-x_{h'})^2}
		\end{equation}
		\begin{equation}
		\Gamma_{\phi'\to WW}
		=
		\frac{9\xi_\varphi^2 \xi_h^2 \kappa^4}{4\pi}
		m_{\phi'}^3
		\frac{\sqrt{1-4x_W}(1-4x_W+12x_W^2)}{(1-x_{h'})^2}
		\end{equation}
		\begin{equation}
		\Gamma_{\phi'\to ZZ}
		=
		\frac{9\xi_\varphi^2 \xi_h^2 \kappa^4}{8\pi}
		m_{\phi'}^3
		\frac{\sqrt{1-4x_Z}(1-4x_Z+12x_Z^2)}{(1-x_{h'})^2}
		\end{equation}
		\begin{equation}
		\Gamma_{\phi'\to h'h'}
		=
		\frac{9\xi_\varphi^2 \xi_h^2 \kappa^4}{8\pi}
		m_{\phi'}^3
		\frac{\sqrt{1-4x_{h'}}(1+2x_{h'})^2}{(1-x_{h'})^2}
		\end{equation}
where $x_a=m_a^2/m_{\phi'}^2$, and $N^{(f)}=3$ is the number of colors of quarks; $N^{(f)}=1$ for leptons. Approximate expressions of three-body decays could be obtained from the large dark matter mass limit, $m_{\phi'}\gg m_{\text{SM}}$, where $m_{\text{SM}}$ is the mass of the heaviest SM particles. The results are:
		\begin{equation}
		\Gamma_{\phi'\to WWh'}
		\simeq
		\frac{3\xi_\varphi^2 \xi_h^2 \kappa^4}{2 (8\pi)^3 } \frac{m_{\phi'}^5}{v^2}
		\end{equation}
		\begin{equation}
		\Gamma_{\phi'\to ZZh'}
		\simeq
		\frac{3\xi_\varphi^2 \xi_h^2 \kappa^4}{4(8\pi)^3 } \frac{m_{\phi'}^5}{v^2}
		\end{equation}
		\begin{equation}
		\Gamma_{\phi'\to h'h'h'}
		\simeq
		\frac{27\xi_\varphi^2 \xi_h^2 \kappa^4}{4(4\pi)^3 } \frac{m_{\phi'} m_{h'}^4}{v^2}
		\end{equation}
	\end{subequations}
	The factor $\xi_h^2$, common to all expressions, indicates that the decay rates could be significantly enhanced by nonminimal coupling between the Higgs field and gravity. This fact is particularly intriguing because it enhanced the possibility of detecting signals arising from decaying scalar singlet dark matter.

\subsection{Compare AD channels with MD channels}

\begin{figure}[!htb]
	\subfigure[]{\includegraphics[width=0.32\textwidth]{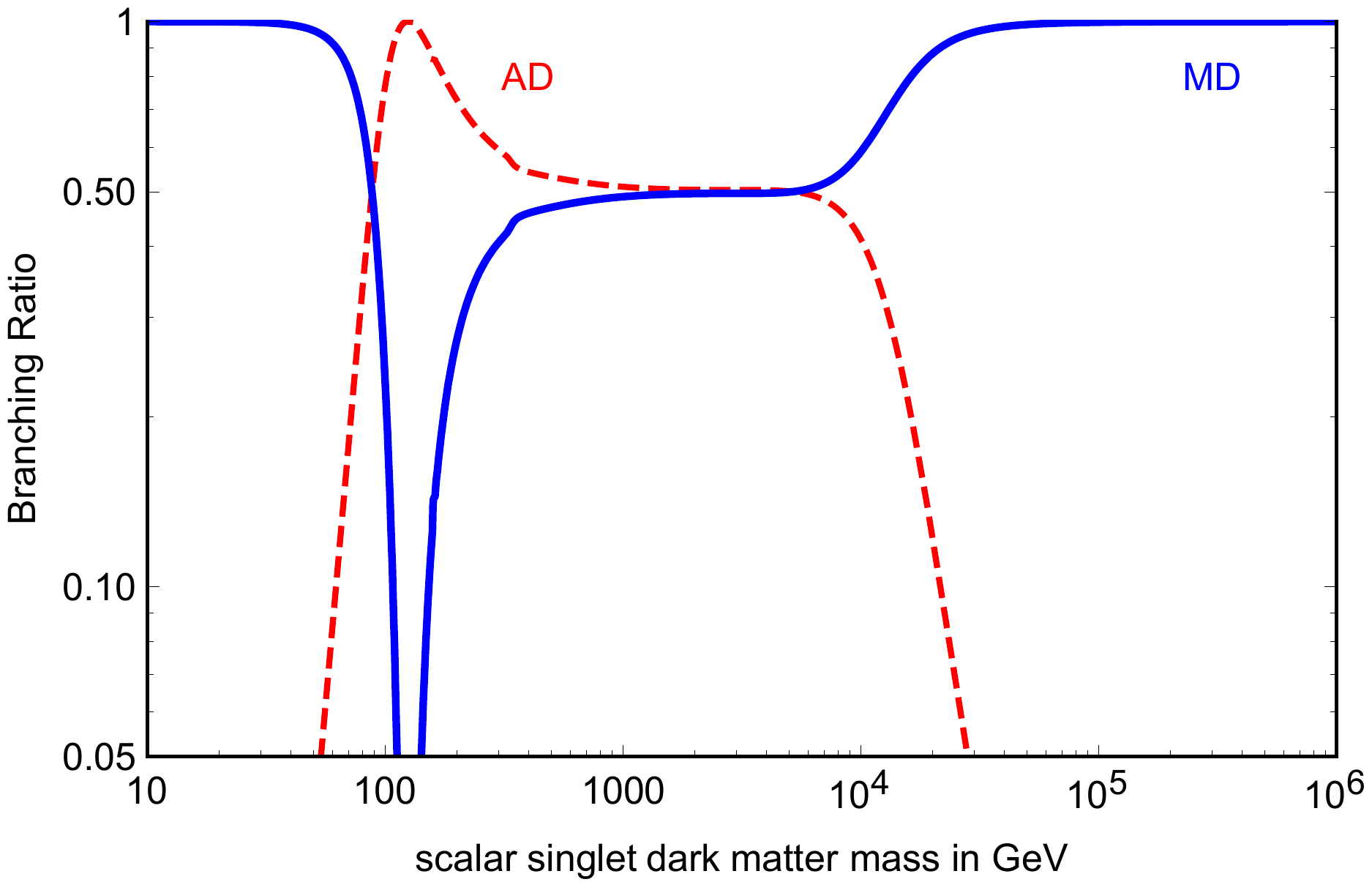}} 
	\subfigure[]{\includegraphics[width=0.32\textwidth]{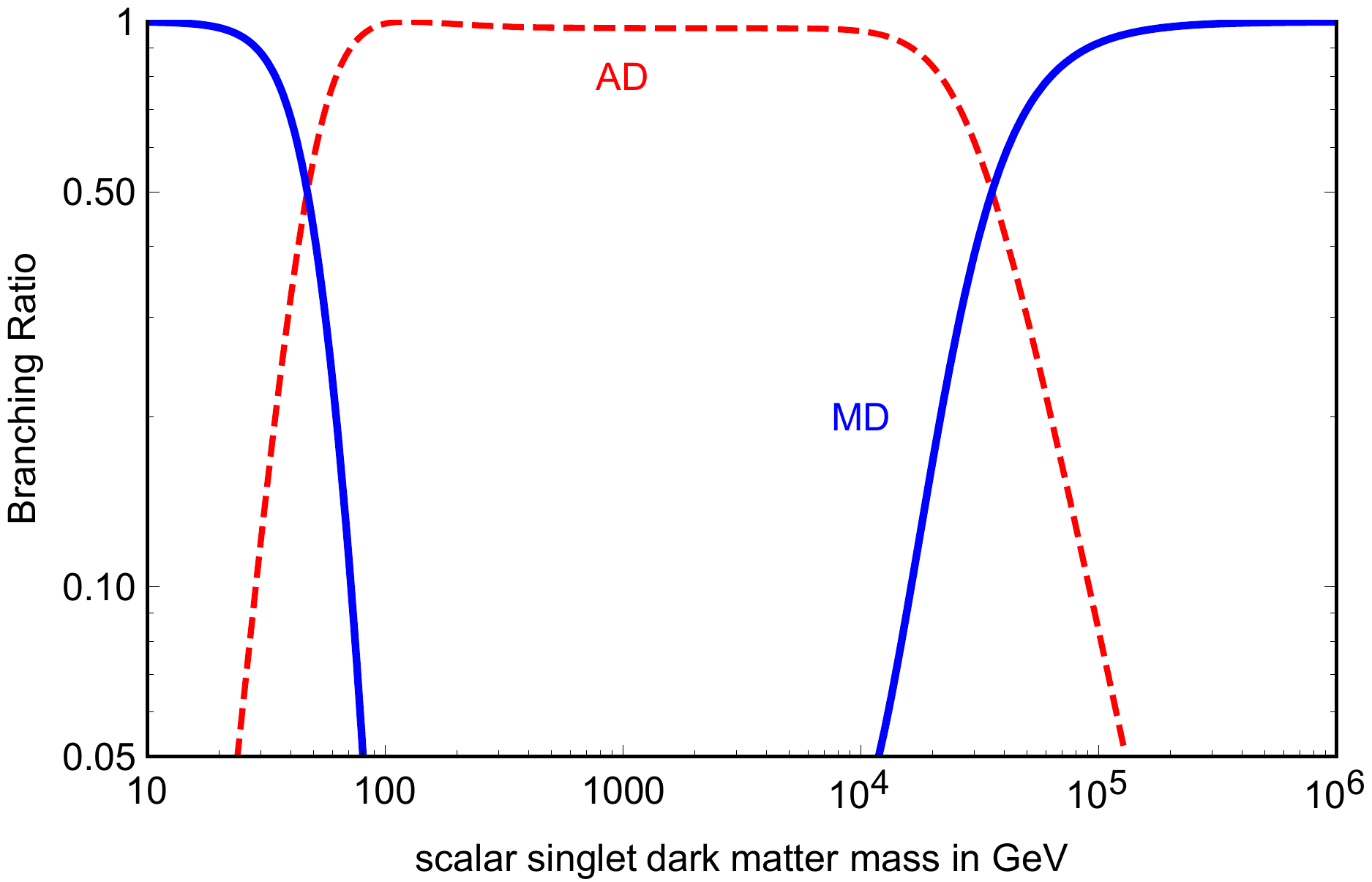}} 
	\subfigure[]{\includegraphics[width=0.32\textwidth]{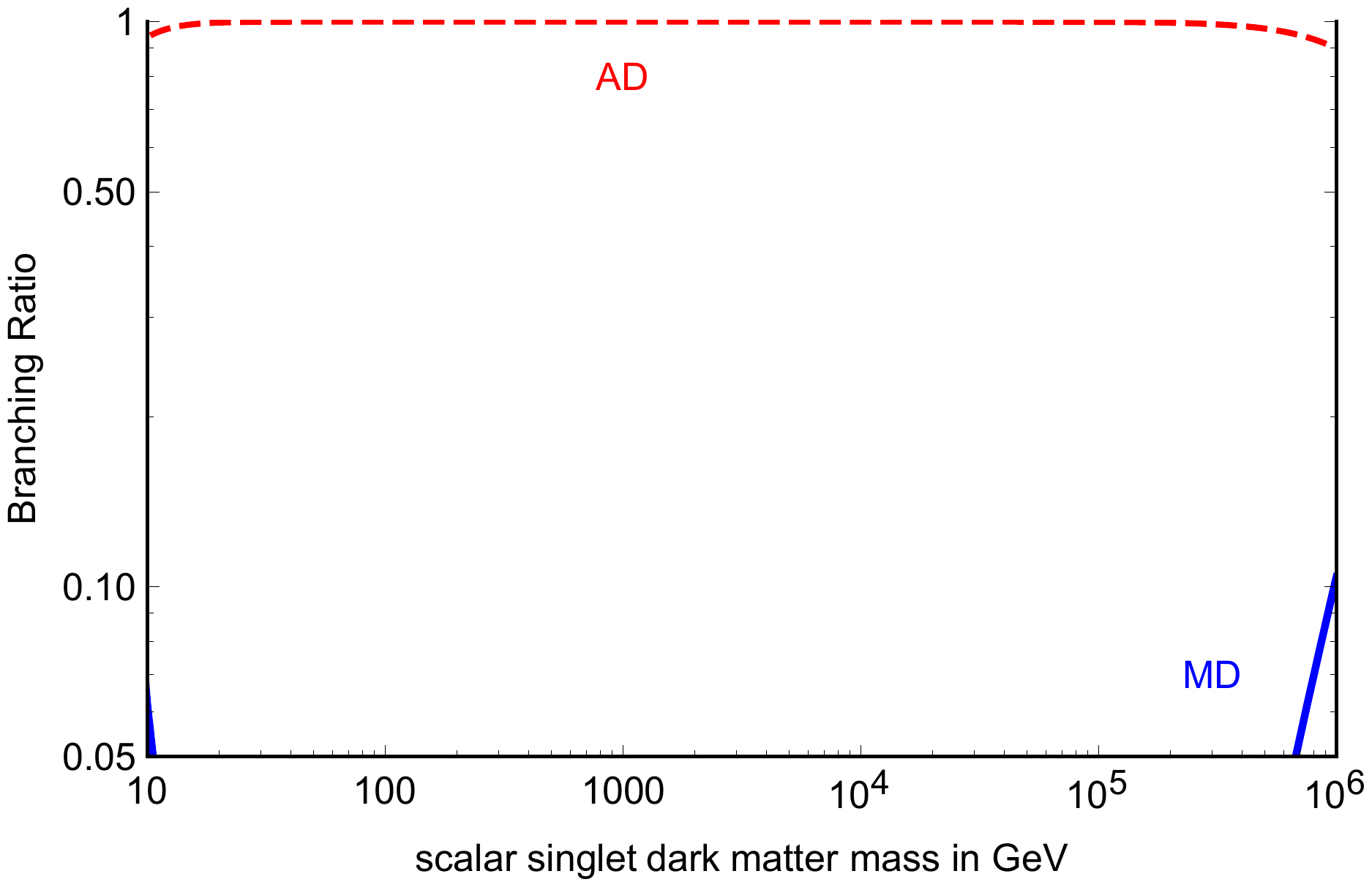}} 
	\caption{\label{fig:BRnewVSoldScalar}
		The branch ratios of total MD channels and total AD channels respectively.\\
		 (a) $\xi_h=1/6$(the conformally coupled case); (b) $\xi_h=1$; (c) $\xi_h=100$;
	}
\end{figure}

Note that when the mass of dark matter is less than 5 TeV, both the gravity and gravity-Higgs portals are dominated by two-body decay channels. The branching ratios of the two-body decay channels in the MD scenario shown by Fig.~\ref{fig:correctedBR} (a) are almost the same as that in the AD scenario shown in Fig.~\ref{fig:BR}. Quantitatively,
\begin{equation}
\frac{\Gamma_{\phi'\to X,X}^\text{AD}}{\Gamma_{\phi\to X,X}^\text{MD}}
=(6\xi_h)^2 (1-x_{h'})^{-2}
\label{eq:proportional}
\end{equation}
where $X,X$ represents a pair of Higgs bosons, W bosons, Z bosons, or fermions. This proportional relationship is independent of specific two-body decay channels. So this proportional relationship reveals that if the mass of dark matter is smaller than 5 TeV, we can not easily judge whether dark matter decays from gravity portals (MD channels) or gravity-Higgs portals (AD channels) from the observation of the decay products. There is an explanation for the proportional relationship expressed by Eq.~\ref{eq:proportional}. The fermions, W bosons, Z bosons, and the Higgs bosons gain their masses through the Higgs mechanism, so all the two-body MD rates are associated with the vacuum expectation value $v$. The gravity-Higgs portal bridges dark matter and the fermions, W bosons, Z bosons, and the Higgs bosons. So all the two-body AD channels are associated with the messenger $h$. Therefore, it is understandable that MD and AD channels' branching ratios are similar.

It is of particular interest that when $\xi_h=1/6$, the Higgs field and gravity are conformally coupled~\cite{Bire82}.
Fig.~\ref{fig:BRnewVSoldScalar} (a) shows the branch ratios of total AD channels and total MD channels of the conformally coupled case. In the case of low mass dark matter, the small rotation angle $\theta \simeq \alpha x_h(1-x_h^2)^{-1}$ suppresses the AD channel $\phi'\to f \bar{f}$, so MD channels are dominant. When the dark matter mass is large enough, the four-body decay in the MD scenario has a large final phase space, which greatly improves the MD rate. Except for dark matter with a mass around 125 GeV, where the sharp increase of the rotation angle leads to the sharp increase of AD rate, the MD and AD rates are almost equivalent between hundreds of GeV and 10 TeV.

From the trend of Fig.~\ref{fig:BRnewVSoldScalar} (a) to (c), with the increase of $\xi_h$, AD channels will gradually dominate in all dark matter mass ranges. In particular, as shown in Fig.~\ref{fig:BRnewVSoldScalar} (c), when $\xi_h=100$, AD channels dominates the decay of dark matter with mass from 10 GeV to $10^6$ GeV. It is noted that the LHC observations only constrained $\xi_h$ to less than $2.6\times10^{15}$~\cite{BoundsHiggsGravity}, 100 is a fairly small coupling constant. Therefore, the possibility remains that gravity-Higgs portals dominate the decay of scalar gravitational dark matter.

\subsection{Observational constraints}

	\begin{figure}[htbp]
	\includegraphics[scale=0.7]{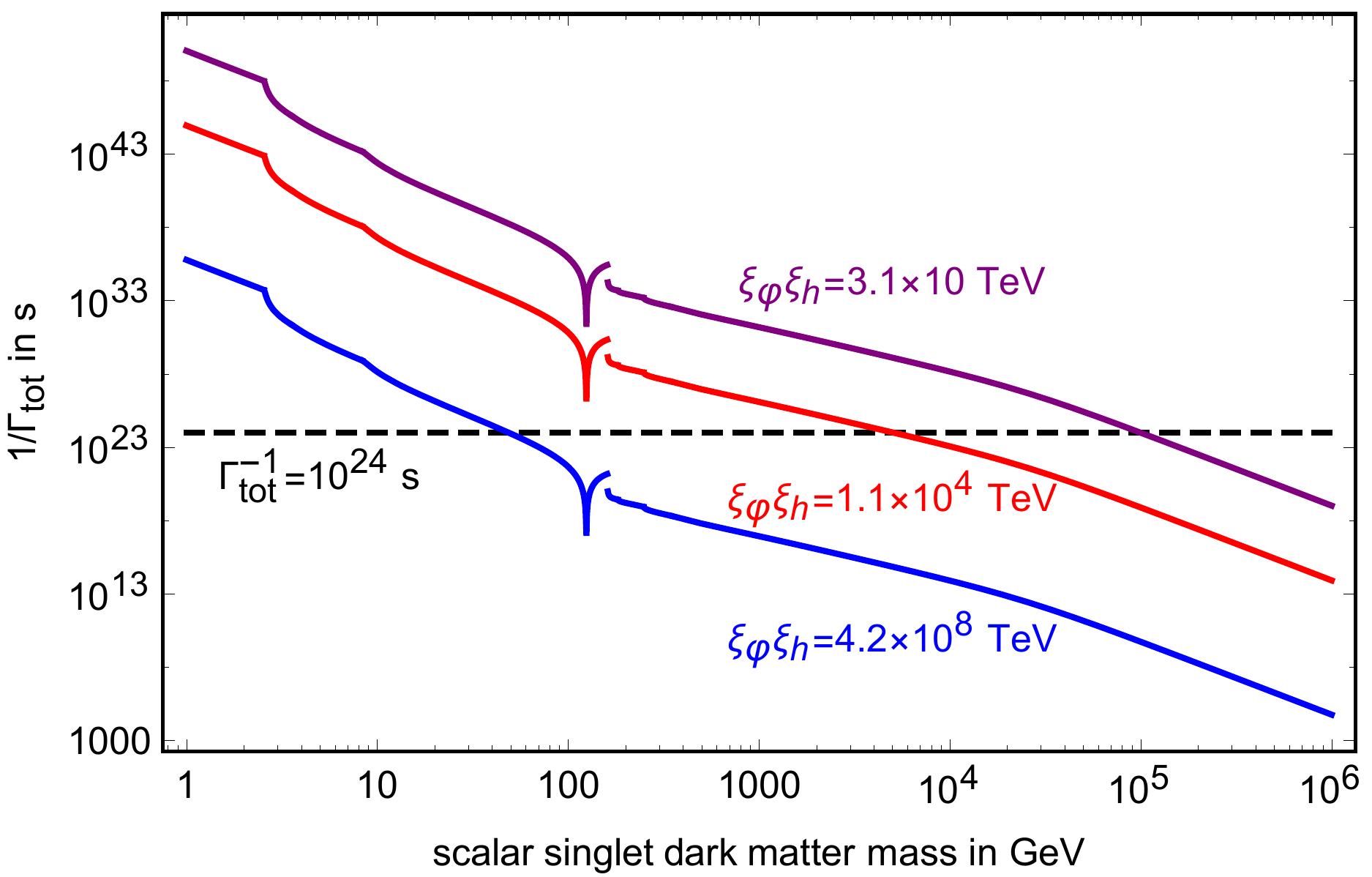}
	\caption{\label{fig:InverseDecayWidth} Inverse decay width as a function of the scalar singlet dark matter mass for $\xi_\varphi\xi_h=4.2\times10^{8}~\text{TeV}, 1.1\times10^{4}~\text{TeV}$, and $3.1\times10~\text{TeV}$. For comparison, conservative constraints from observations, e.g. $\Gamma^{-1}>10^{24}~\text{s}$, are also shown.}
\end{figure}

The decay products of dark matter could contribute to the observed cosmic abundances of $\gamma$-rays, neutrinos, and charged particles. The contributed fluxes by two-body decay channels are supplied by PPPC 4 DM ID~\citep{cirelli2011} and~\citet{updateToThePPPC4DM}. It is suggested that the total amount of decay products cannot exceed the observed fluxes. Along this line, many inspiring works have derived lower limits (upper limits) to the lifetime (decay rates) of dark matter. Employing the isotropic diffuse $\gamma$-ray background reported by Fermi-LAT~\citep{FermiLAT}, stringent limits on a wide range of two-body decay channels and masses (from GeV to EeV and above) was derived by~\citet{CfIGRB}, which were generally in the range of $\Gamma^{-1}\sim(1-5)\times10^{28}~\text{s}$. Positron data and antiproton data observed by AMS-02 and cosmic neutrino flux observed by neutrino telescopes such as IceCube, Auger experiments are also used to set constraints on the lifetime of dark matter candidates~\citep{PhysRevD89063539,PhysRevLett113121101,PhysRevLett117091103,Esmaili2012}. Conservative limits on the coexistence of many two-body and three-body channels were derived by~\citet{SunandDai} who found that the lower limit to the lifetime of dark matter is $10^{26}~\text{s}$. Unfortunately, limits on $\phi'\to WWh',ZZh'$ channels have been rarely studied. Given these considerations, this paper uses the conservative condition of $\Gamma^{-1}>10^{24}~\text{s}$ to set conservative constraints on the coupling constants. It is noteworthy that this conservative constraint is more stringent than the constraint from the age of the universe, $\Gamma^{-1}>4\times 10^{17}~\text{s}$.

	Figure~\ref{fig:InverseDecayWidth} show the total inverse decay widths for $\xi_\varphi\xi_h=4.2\times10^{8}~\text{TeV}, 1.1\times10^{4}~\text{TeV}$, and $3.1\times10~\text{TeV}$. The conservative lower limit from observations of $\Gamma^{-1}>10^{24}~\text{s}$ is also plotted for comparison. It is apparent that scalar singlet dark matter with heavier masses receive more stringent constraints from the observations. Specifically, the blue curve shows that $|\xi_\varphi\xi_h|>4.2\times10^{8}~\text{TeV}$ can be excluded when the mass of dark matter is 50~GeV, the red curve shows that $|\xi_\varphi\xi_h|>1.1\times10^{4}~\text{TeV}$ can be excluded when the mass of dark matter is 5~TeV, and the purple curve shows that $|\xi_\varphi\xi_h|>3.1\times10~\text{TeV}$ can be excluded when the mass of dark matter is 100~TeV.

More concretely, approximate lower lifetime bounds for many interesting mass regions can also be obtained. As analyzed previously, the total decay rate is dominated by different channels for different dark matter masses. The dominant decay channel is $\phi'\to b\bar{b}$ for $m_{\phi'}\sim 10-160~\text{GeV}$, $\phi'\to WW,ZZ,h'h'$ for $m_{\phi'}\sim 160~\text{GeV}-20~\text{TeV}$, and $\phi'\to WWh',ZZh'$ for $m_{\phi'}\sim 20~\text{TeV}-10^3~\text{TeV}$. We have used the dominant channels to estimate the approximated decay rate and some rough constraints on nonminimal coupling constants. In the limit of large dark matter mass, the approximated decay rates are:
		\begin{equation}
		\Gamma_{\text{tot}}
		\simeq
		\frac{9\xi_\varphi^2 \xi_h^2 \kappa^4}{2\pi}
		m_{\phi'}^3
		\times
		\left\{
		\begin{aligned}
		3 x_b
		\frac{(1-4x_b)^{3/2}}{(1-x_h)^2}
		,&&m_{\phi'}\sim 10-160~\text{GeV} \\
		1,&&m_{\phi'}\sim 160~\text{GeV}-20~\text{TeV} \\
		\frac{m_{\phi'}^2}{(32 \pi v)^2 } ,&&m_{\phi'}\sim 20~\text{TeV}-10^3~\text{TeV}
		\end{aligned}
		\right.
		\end{equation}
where the subscript $b$ in $x_b$ represents the bottom quark. Using conditions that $\Gamma^{-1}>10^{24}~\text{s}$, these expressions can be translated into upper bounds on the nonminimal coupling parameters as:
\begin{equation}
|\xi_\varphi \xi_h|
\stackrel{<}{\sim}
\left\{
\begin{aligned}
4.2\times10^{8}
~\text{TeV}
,&&m_{\phi'}\sim 50~\text{GeV} \\
1.1\times10^{4}
(\frac{m_{\phi'}}{5~\text{TeV}})^{-1.5}
~\text{TeV}
,&&m_{\phi'}\sim 160~\text{GeV}-20~\text{TeV} \\
3.1\times10
(\frac{m_{\phi'}}{100~\text{TeV}})^{-2.5}
~\text{TeV}
,&&m_{\phi'}\sim 20~\text{TeV}-10^3~\text{TeV}
\end{aligned}
\right.
\label{eq:ssDMconstraints}
\end{equation}
The obtained limits are more stringent when the mass of the scalar singlet dark matter is heavier. The results also show that when we consider TeV physics, there is still a vast parameter space alvie for TeV scalar singlet dark matter.

\section{Fermionic singlet as dark matter candidate}\label{FermionicDMdecay}

Fermionic singlet dark matter (denoted by $\chi$) also have been extensively studied in the literature~\citep{fermionicsinglet}. It is also assumed to be charged under a discrete $\mathbb{Z}_2$ symmetry, under which $\chi$ is odd and all SM fields are even. For the fermionic singlet dark matter, it could checked that $F(\varphi, X)R=(\bar{l}\breve{H}\chi+\bar{\chi}\breve{H}^\dag l)R$ is the operator of the lowest dimension that can break the hypothetical $\mathbb{Z}_2$ symmetry~\citep{DMdecayTGP}, where $\breve{H}=i\sigma_2 H^*$, $l$ stands for electroweak lepton doublet, $\boldsymbol{\sigma}=(\sigma_1,\sigma_2,\sigma_3)$ are the Pauli matrices.

	\begin{table*}[htbp]
		\caption{\label{vertexrules2}
			Feynman rules for the decay of fermionic singlet dark matter.
			\footnote{In the Feynman rules, the symmetry factors are taken into account.}
		}
		\begin{ruledtabular}
			\begin{tabular}{lcdr}
				\textrm{Physical process}&
				\textrm{Feynman rules\footnote{$p_\chi,p_h,p_{\nu_l}$ denotes the four momenta of the dark matter particles, Higgs boson, and neutrinos, respectively.}}\\
				\colrule
				$\chi\to \nu_l  h$
				&  $
				i3\sqrt{2}\kappa^2
				\xi_h\xi_\varphi v^2
				(p_\chi p_h-p_{\nu_l} p_h)/2
				$   \\
				$\chi\to \nu_l h h$
				&  $
				i 3\sqrt{2}\kappa^2
				\xi_h\xi_\varphi  v
				( p_\chi p_{h1}+p_\chi p_{h2}-p_{\nu_l} p_{h1}- p_{\nu_l} p_{h2}-p_{h1} p_{h2})
				$   \\
				$\chi\to \nu_l h h h$
				&   $
				i3\sqrt{2}\kappa^2
				\xi_h\xi_\varphi
				[\sum_{\text{j}=1}^{3} (p_\chi p_{h\text{j}} - p_{\nu_l} p_{h\text{j}}) -p_{h1} p_{h2}-p_{h1} p_{h3}-p_{h2} p_{h3}]
				$  \\
			\end{tabular}
		\end{ruledtabular}
	\end{table*}
	
	\begin{table}[htbp]
		\caption{\label{decaymodes2}
			Tree-level decay modes of the fermionic singlet dark matter.
		}
		\begin{ruledtabular}
			\begin{tabular}{lcdr}
				\textrm{Decay mode}&
				\textrm{Asymptotic scaling}\\
				\colrule
				$\chi\to \nu_l h$
				& $m_\chi v^8$ \\
				$\chi\to \nu_l hh$
				& $m_\chi^7 v^2$ \\
				$\chi\to \nu_l hhh$
				& $m_\chi^{9} $ \\
			\end{tabular}
		\end{ruledtabular}
	\end{table}
	
	\begin{figure}[htbp]
		\includegraphics[scale=0.7]{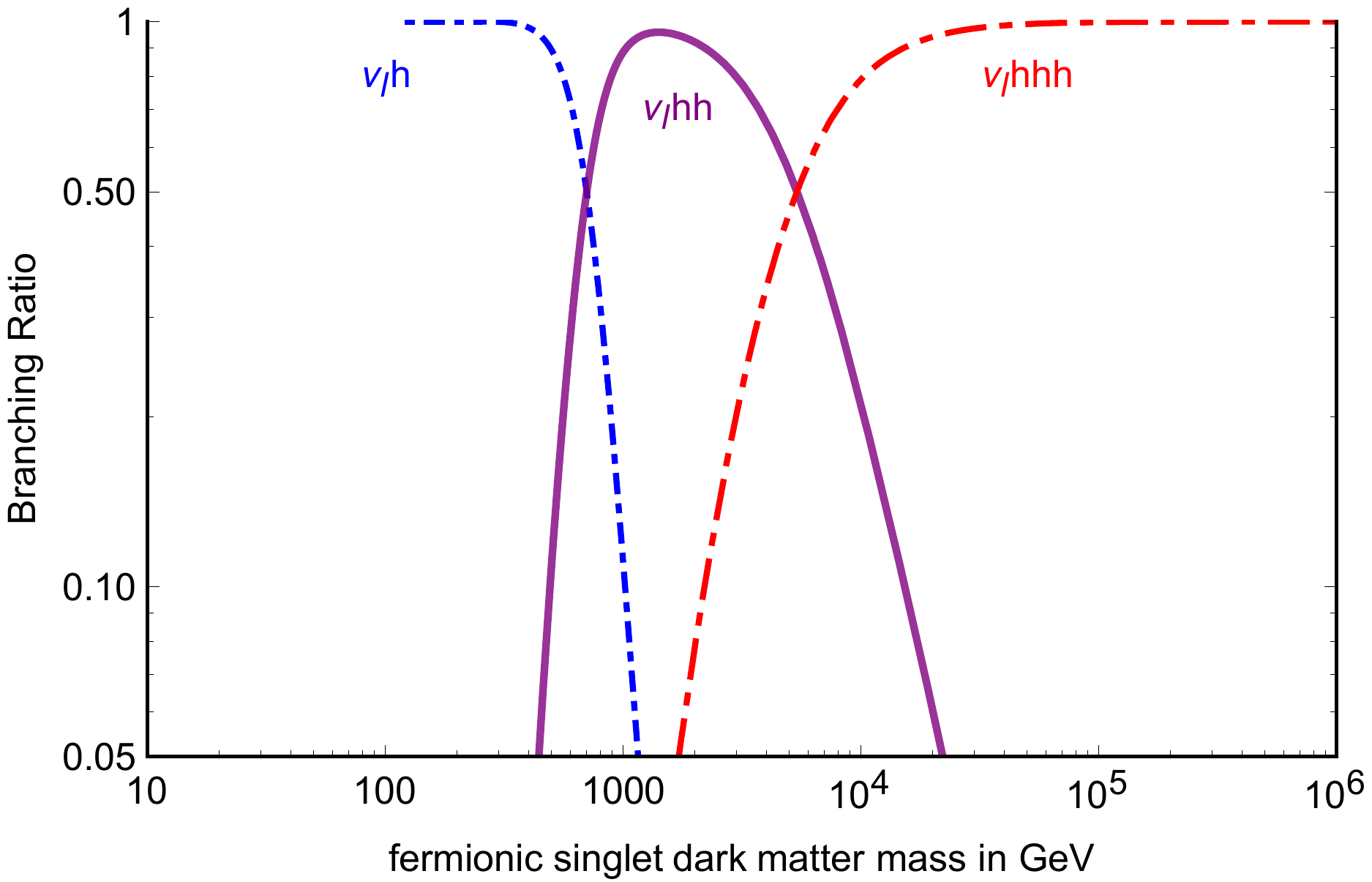}
		\caption{\label{fig:BR2} Decay branching ratios for the fermionic singlet dark matter as a function of its mass.}
	\end{figure}

\subsection{Branch ratios}
	Similar to the previous situation, let us focus on the effective interactions $3\kappa^{-2} \Omega_{,\rho}\tilde{\Omega}^{,\rho} \Omega^{-2} $. Working with the unitary gauge, after electroweak symmetry breaking, the Weyl factor becomes $\Omega^2=1+\kappa^2 \xi_h (v+h)^2 +\sqrt{2}\kappa^2\xi_\varphi(v+h)(\bar{\nu}_l\chi+\bar{\chi}\nu_l)$, where $\nu_l$ stands for neutrino. Then, I only write out the terms in the effective interactions relating to the decay of the fermionic singlet:
	\begin{eqnarray}
	\frac{3\kappa^2}{4\Omega}
	2\sqrt{2}\xi_h\xi_\varphi v^2
	(\bar{\nu}_l \chi^{,\rho}+\bar{\nu}_l^{,\rho} \chi) h_{,\rho}
	+
	\frac{3\kappa^2}{4\Omega}
	2\sqrt{2}\xi_h\xi_\varphi v h
	(\bar{\nu}_l \chi^{,\rho}+\bar{\nu}_l^{,\rho} \chi) h_{,\rho}
	+
	\frac{3\kappa^2}{4\Omega}
	2\sqrt{2}\xi_h\xi_\varphi  v       \nonumber\\
	\times
	(\bar{\nu}_l \chi^{,\rho} h+\bar{\nu}_l^{,\rho} \chi h+\bar{\nu}_l \chi h^{,\rho}) h_{,\rho}
	+
	\frac{3\kappa^2}{4\Omega}
	2\sqrt{2}\xi_h\xi_\varphi  h
	(\bar{\nu}_l \chi^{,\rho} h+\bar{\nu}_l^{,\rho} \chi h+\bar{\nu}_l \chi h^{,\rho}) h_{,\rho}
	\end{eqnarray}
	note that the fermionic fields $\nu_l$ and $\chi$ are rescaled through Eq.~\ref{eq:rescalefermions}.
	Different from the scalar singlet case, no kinetic mixing term appears here. Hence, the fermionic singlets could only decay into Higgs bosons and neutrinos. For this case, Table~\ref{vertexrules2} has provided Feynman rules for the decay of fermionic singlet dark matter. In Table~\ref{decaymodes2}, all possible tree-level decay channels are listed, together with the asymptotic scaling of large dark matter masses. Figure~\ref{fig:BR2} shows the exact decay branch ratios of the fermionic singlet dark matter. It can be seen that the decay phenomenology of the fermionic singlet is quite different from the scalar singlet case.
	
	There are only three channels for a fermionic singlet to decay, $\chi\to \nu_l h$, $\chi\to \nu_l hh$, and $\chi\to \nu_l hhh$. Thus, when the mass of the fermionic singlet is below $m_h\sim 125~\text{GeV}$, the decay width does not increase even if the nonminimal coupling term between the Higgs field and gravity exists.
		
	For fermionic singlet masses between $m_h$ and 700 GeV, the decay rate is dominated by $\chi\to \nu_l h$, since the $\chi\to \nu_l hh$ channel is suppressed by the small phase space. Specifically, when the mass of the fermionic singlet is between $m_h$ and $2m_h$, it is the only channel allowed by the phase space. For intermediate masses, $700~\text{GeV}\leq m_\chi\leq 5.4~\text{TeV}$, the decay rate is dominated by channel $\chi\to \nu_l hh$. In this range, compared with $\chi\to \nu_l hh$, $\chi\to \nu_l h$ is suppressed by a factor of $v^6/m_\chi^6$. Meanwhile, the small phase space suppressed $\chi\to \nu_l hhh$. For large fermionic singlet masses above 5.4 TeV, $\chi\to \nu_l hhh$ dominates the decay rate due to the large asymptotic scaling.
	
	It can be concluded that the decay phenomenology of the fermionic singlet is quite different from the scalar singlet case. The reason for this is that the electroweak symmetry breaking could result in a kinetic mixing term between the scalar dark matter and the Higgs field, consequently resulting in the decay of scalar dark matter into SM particles through the Higgs field. Conversely, fermionic dark matter cannot kinetically mix with the Higgs field, so it mainly decays into coupled particles, namely, neutrinos and Higgs bosons.
	
	The analytical expressions of the decay rates of the two-body and three-body decay channels are:
	\begin{subequations}
		\begin{equation}
		\Gamma_{\chi\to\nu_l h}
		=
		\frac{9\xi_\varphi^2 \xi_h^2 \kappa^4}{64\pi} m_\chi m_h^4 v^4 (1-x_h)^2
		\end{equation}
		\begin{equation}
		\Gamma_{\chi\to\nu_l hh}
		=
		\frac{3\xi_\varphi^2 \xi_h^2 \kappa^4}{5(16\pi)^3} m_\chi^7 v^2 g(x_h)
		\end{equation}
		with
		\begin{eqnarray}
		g(x)=
		(1+7x-44x^2+810x^3-1260x^4)\sqrt{1-4x}
		+
		240x^3(5-17x+21x^2)\text{log}(\frac{2\sqrt{x}}{1+\sqrt{1-4x}}) \nonumber
		\end{eqnarray}
where $x_h=m_h^2/m_\chi^2$. Approximate expressions of the four-body decays could be obtained in the large dark matter mass limit, $m_\chi\gg m_h$ as:
		\begin{equation}
		\Gamma_{\chi\to\nu_l hhh}
		\simeq
		\frac{9\xi_\varphi^2 \xi_h^2 \kappa^4}{80(8\pi)^5} m_\chi^9
		\end{equation}
	\end{subequations}
       These expressions also show that nonminimal coupling between the Higgs field and gravity can also enhance the possibility of detecting signals from decaying fermionic singlet dark matter.

\subsection{Compare AD channels with MD channels}

\begin{figure}[!htb]
	\subfigure[]{\includegraphics[width=0.32\textwidth]{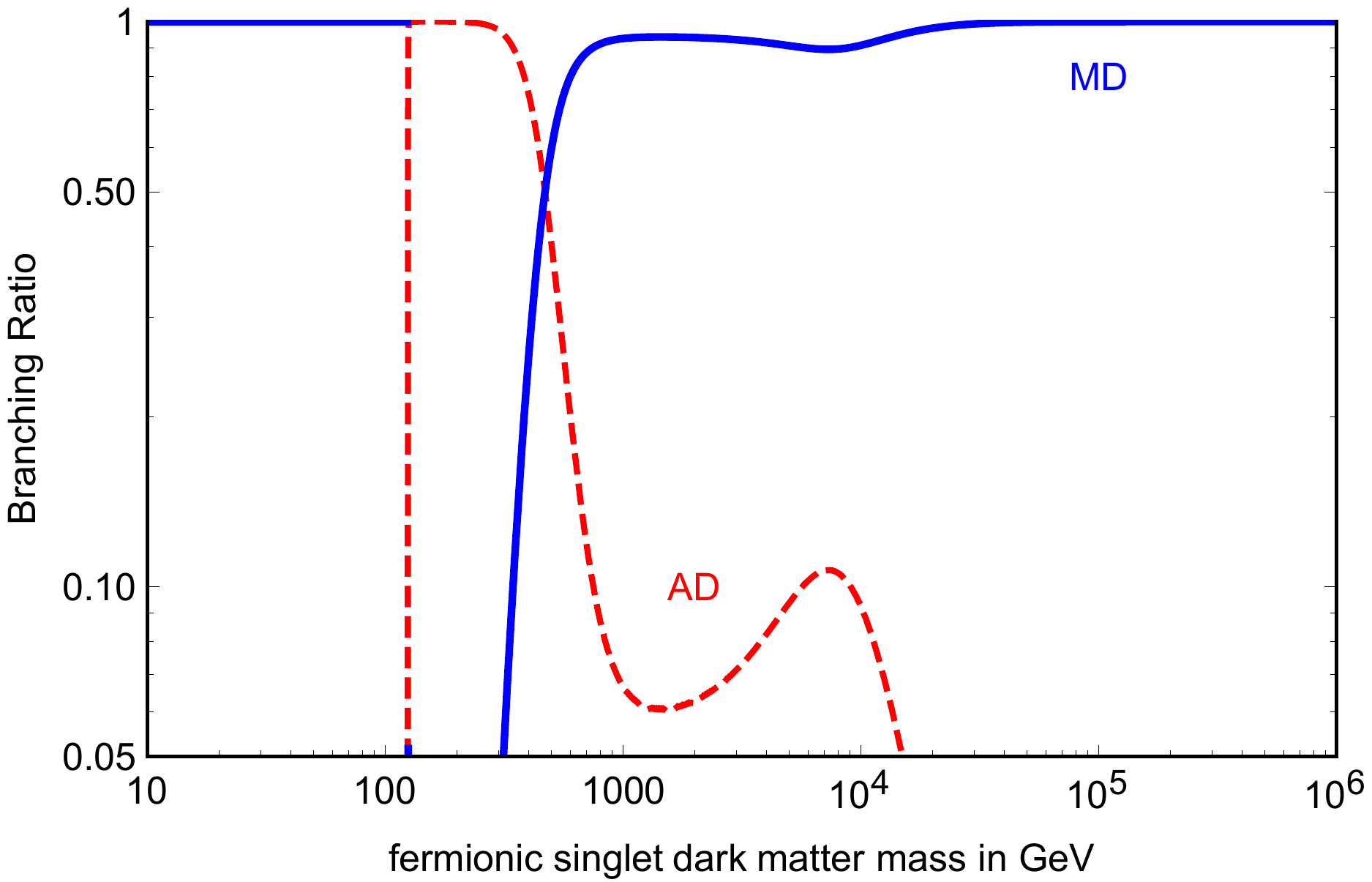}} 
	\subfigure[]{\includegraphics[width=0.32\textwidth]{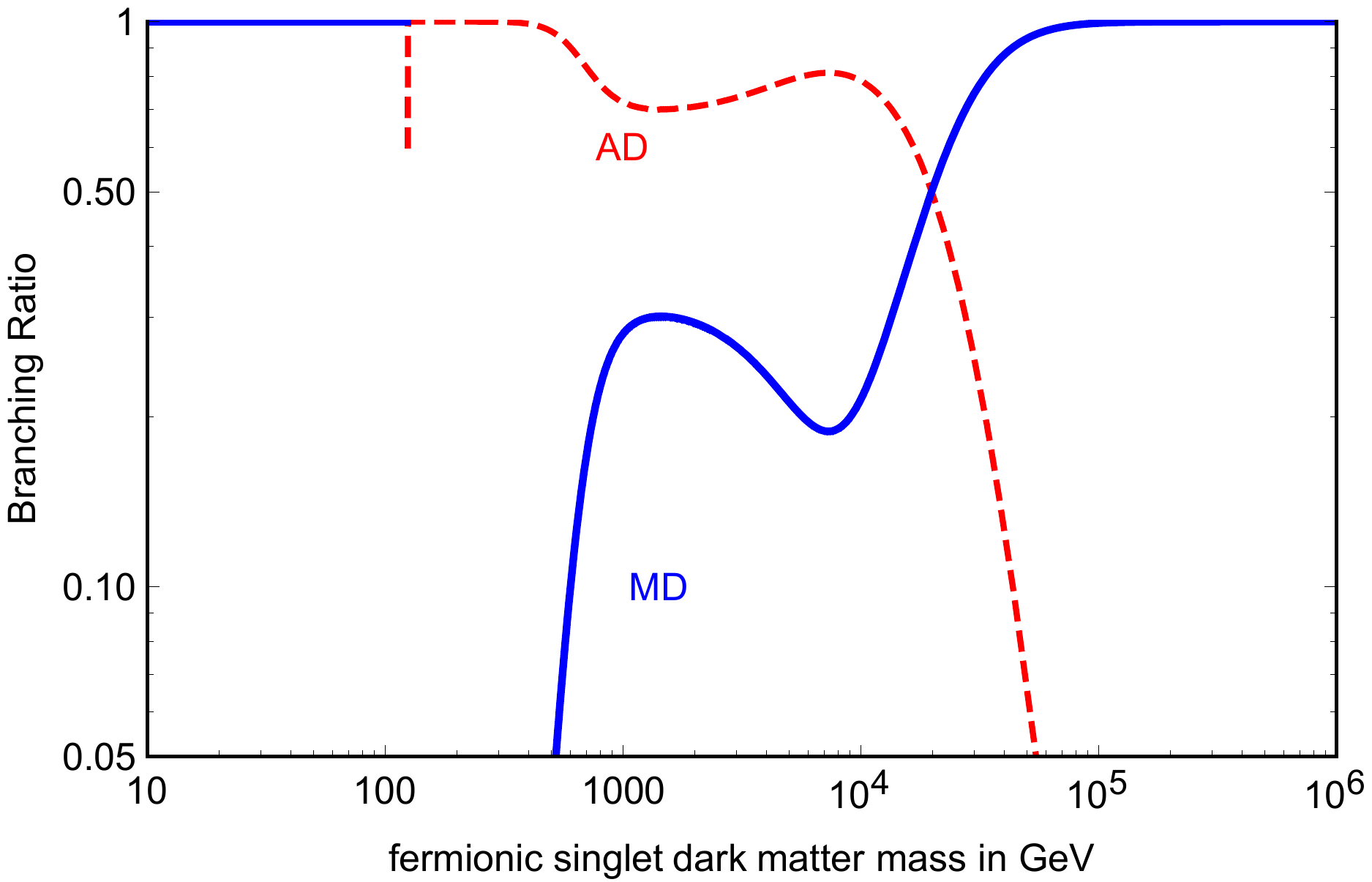}} 
	\subfigure[]{\includegraphics[width=0.32\textwidth]{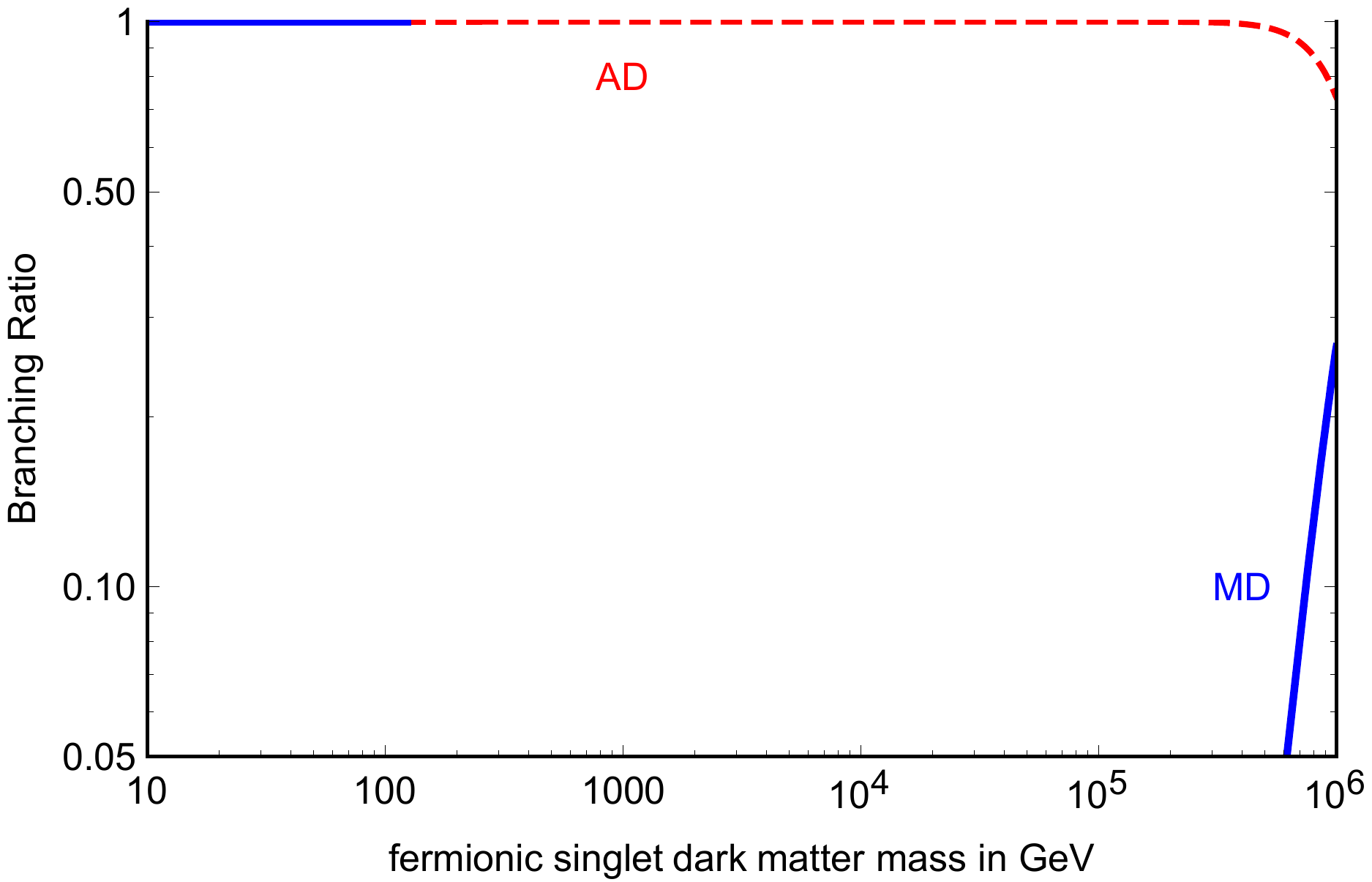}} 
	\caption{\label{fig:BRnewVSoldFermionic}
		The branch ratios of total MD channels and total AD channels respectively.\\
		(a) $\xi_h=1/6$(the conformally coupled case); (b) $\xi_h=1$; (c) $\xi_h=2000$;
	}
\end{figure}

	Fig.~\ref{fig:BRnewVSoldFermionic} (b) shows when $\xi_h=1$, the branch ratios of total AD channels and total MD channels. As for the low-mass dark matter, AD channels are kinematically forbidden, and MD channels are dominant. When the mass of dark matter is between 125 GeV and 500 GeV, MD channels are greatly suppressed by AD channels (specifically, $\chi\to h \nu$ channel). When the mass of dark matter is in the range of 500 GeV$\sim$10 TeV, the MD and AD rates are in the same order of magnitude. When the dark matter mass is large enough, the $\chi\to WWhh\nu+ZZhh\nu$ and $\chi\to WWhhh\nu+ZZhhh\nu$ channels in the MD scenario has a large final phase space, which greatly improves the MD rate.

	From the trend of figures~\ref{fig:BRnewVSoldFermionic} (a) to (c), with the increase of $\xi_h$, AD channels will gradually dominate in the range of $m_\chi>m_h$. In particular, as shown in Fig.~\ref{fig:BRnewVSoldFermionic} (c), when $\xi_h=2000$, AD channels dominates the decay of dark matter with mass from 125 GeV to $10^6$ GeV. It is noted that the LHC observations only constrained $\xi_h$ to less than $2.6\times10^{15}$~\cite{BoundsHiggsGravity}, 2000 is a fairly small coupling constant. Therefore, the possibility remains that gravity-Higgs portals dominate the decay of fermionic gravitational dark matter.

\subsection{Observational constraints}

	\begin{figure}[htbp]
	\includegraphics[scale=0.7]{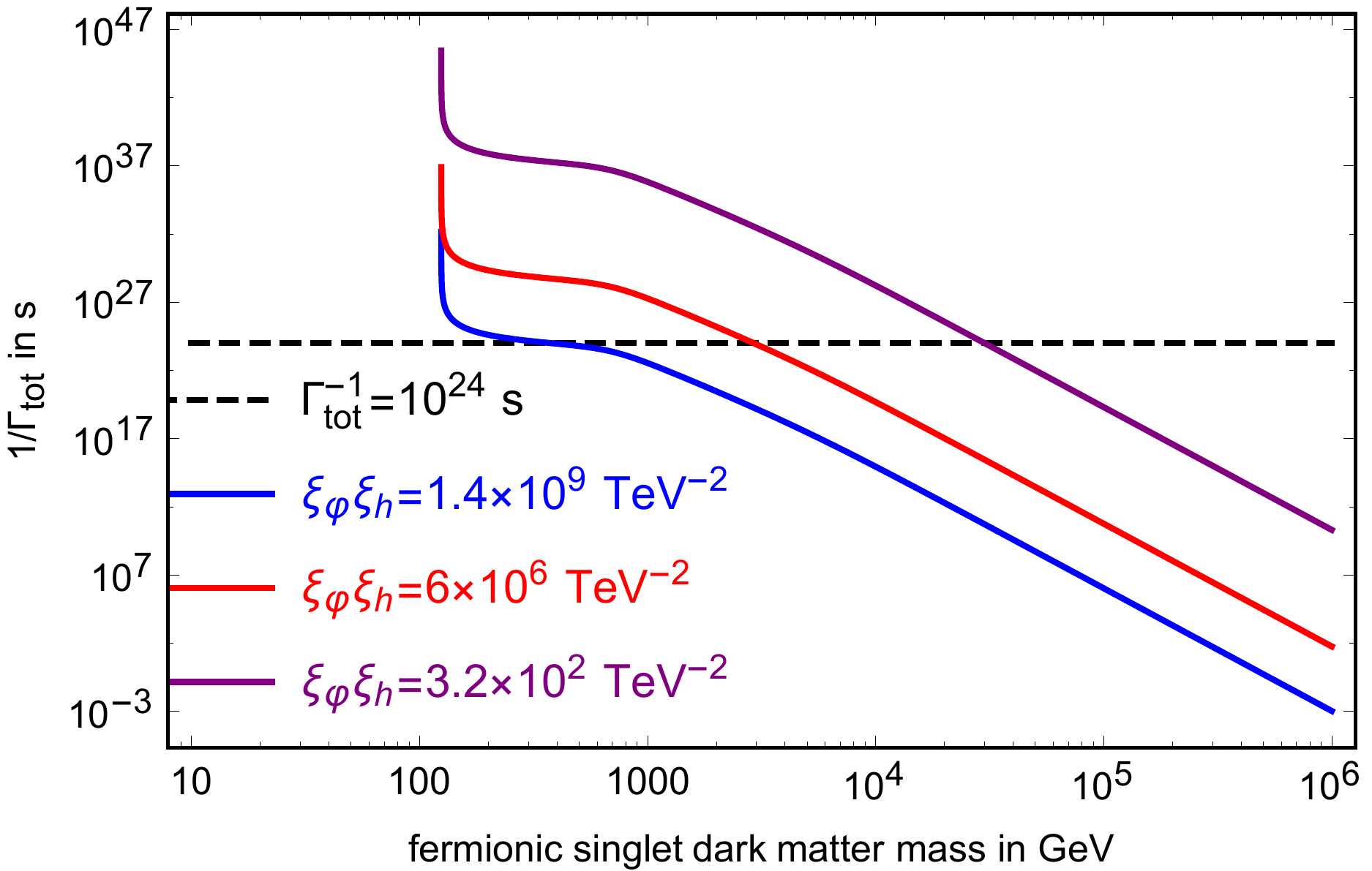}
	\caption{\label{fig:InverseDecayWidth2} Inverse decay width as a function of the fermionic singlet dark matter mass for $\xi_\varphi\xi_h=1.4\times10^9~\text{TeV}^{-2}, 6\times10^6~\text{TeV}^{-2}$, and $3.2\times10^{2}~\text{TeV}^{-2}$. For comparison, the conservative constraint from observations of $\Gamma^{-1}>10^{24}~\text{s}$ is also shown.}
\end{figure}

Similar to the scalar singlet case, I determined limits on the coupling constants using the condition that $\Gamma^{-1}>10^{24}~\text{s}$. Fig.~\ref{fig:InverseDecayWidth2} shows the total inverse decay widths for $\xi_\varphi\xi_h=1.4\times10^9~\text{TeV}^{-2}, 6\times10^6~\text{TeV}^{-2}$, and $3.2\times10^{2}~\text{TeV}^{-2}$. The conservative lower limit from observations of $\Gamma^{-1}>10^{24}~\text{s}$ is also plotted for comparison. It can be seen from Fig.~\ref{fig:InverseDecayWidth2} that fermionic singlet dark matter with heavier masses receive more stringent constraints from the observations. Specifically, the blue curve shows that $|\xi_\varphi\xi_h|>1.4\times10^9~\text{TeV}^{-2}$ can be excluded when the mass of dark matter is 300~GeV, the red curve shows that $|\xi_\varphi\xi_h|>6\times10^6~\text{TeV}^{-2}$ can be excluded when the mass of dark matter is 3~TeV, and the purple curve shows that $|\xi_\varphi\xi_h|>3.2\times10^{2}~\text{TeV}^{-2}$ can be excluded when the mass of dark matter is 30~TeV.

Approximate lower bounds of lifetime on the specific mass regions can also be obtained. As analyzed above, the dominant decay channels are $\chi\to \nu_l h$ for $m_\chi\sim m_h-700~\text{GeV}$, $\chi\to \nu_l hh$ for $m_\chi\sim 700~\text{GeV}-5.4~\text{TeV}$, and $\chi\to \nu_l hhh$ for $m_\chi\sim 5.4~\text{TeV}-10^3~\text{TeV}$. I have used these dominant channels to estimate the decay rates and rough constraints on nonminimal coupling constants. The approximated decay rates are:
		\begin{equation}
		\Gamma_{\text{tot}}
		\simeq
		\frac{3\xi_\varphi^2 \xi_h^2 \kappa^4}{64\pi} m_\chi
		\times
		\left\{
		\begin{aligned}
		3 m_h^4 v^4
		,&&m_\chi\sim m_h-700~\text{GeV} \\
		\frac{1}{5(8\pi)^2} m_\chi^6 v^2 ,
		&&m_\chi\sim 700~\text{GeV}-5.4~\text{TeV} \\
		\frac{3}{10(8\pi)^4} m_\chi^8 ,
		&&m_\chi\sim 5.4~\text{TeV}-10^3~\text{TeV}
		\end{aligned}
		\right.
		\end{equation}
		Using conditions that $\Gamma^{-1}>10^{24}~\text{s}$, these expressions could translate into upper bounds on the nonminimal coupling parameters such that:
	\begin{equation}
|\xi_\varphi \xi_h|
\stackrel{<}{\sim}
\left\{
\begin{aligned}
1.4\times10^9
(\frac{m_\chi}{300~\text{GeV}})^{-0.5}
~\text{TeV}^{-2}
,&&m_\chi\sim m_h-700~\text{GeV} \\
6\times10^6
(\frac{m_\chi}{3~\text{TeV}})^{-3.5}
~\text{TeV}^{-2}
,&&m_\chi\sim 700~\text{GeV}-5.4~\text{TeV} \\
3.2\times10^{2}
(\frac{m_\chi}{30~\text{TeV}})^{-4.5}
~\text{TeV}^{-2}
,&&m_\chi\sim 5.4~\text{TeV}-10^3~\text{TeV}
\end{aligned}
\right.
\label{eq:fsDMconstraints}
\end{equation}
		These limits are more stringent when the mass of the fermionic singlet dark matter is heavier. The results also show that when we consider TeV physics, there is still a vast parameter space alvie for TeV fermionic singlet dark matter. Compare Eq.~\ref{eq:ssDMconstraints} with Eq.~\ref{eq:fsDMconstraints} we can conclude that gravity-Higgs portals of scalar singlet dark matter are more constrained than gravity-Higgs portals of fermionic singlet dark matter.

\section{Summary}\label{TheSummary}

Gravity can break the global symmetry that guarantees the stability of dark matter and lead to its decay. This paper studied how nonminimal coupling between the Higgs field and gravity affects the decay of gravitational dark matter. The conclusion is that nonminimal coupling between the Higgs field and gravity can induce an AD decay width of gravitational dark matter.
When $\xi_h=1/6$, the AD width of the scalar singlet dark matter with a mass of $100~\text{GeV}<m_{\phi'}<10~\text{TeV}$ is at least of more than 50\% of total decay width.
When $\xi_h=1$, the AD width of the fermionic singlet dark matter with a mass of $m_h<m_{\chi}<10~\text{TeV}$ is at least of more than 50\% of total decay width.

	In the case of scalar singlet dark matter, this paper finds that if both the Higgs field and the dark matter field couple to the Ricci scalar, respectively, electroweak symmetry breaking can cause kinetic mixing between it and the Higgs field. The kinetic mixing indicates that these two fields are not their mass eigenstates. The kinetic mixing term turns out to be the primary portal for the scalar singlet to decay into SM particles. Then this paper obtained the mass eigenstates of the scalar singlet and Higgs particles through rotation and rescaling operations. The rotation angle turns out to be a key parameter for the decay of the scalar singlet. Next, this study derived Feynman rules involving the main AD channels at the tree level. Finally, the branching ratios of the different decay channels are plotted. For scalar dark matter candidates with a mass less than 5 TeV, the decay phenomenology of gravity portals and gravity-Higgs portals are almost the same. It was revealed that the main AD channel is $\phi'\to f \bar{f}$ for scalar singlets below 160~GeV. The decay channels are abundant for scalar singlets with intermediate masses, specifically, $\phi'\to WW,ZZ,h'h',t\bar{t}$ for $160~\text{GeV}\leq m_{\phi'}\leq 20~\text{TeV}$. Meanwhile, for scalar singlets with heavier masses ($20~\text{TeV}\leq m_{\phi'}\leq 10^3~\text{TeV}$), $\phi'\to WWh',ZZh'$ dominates the AD rate.

	In the case of fermionic singlet dark matter, the decay phenomenology is quite different from the scalar singlet case, as there is no kinetic mixing between it and the Higgs field. Hence, it mainly decays into neutrinos and Higgs bosons. This paper provided Feynman rules involving the main AD channels at the tree level. Then the branching ratios of the gravity-Higgs portals are plotted. It was revealed that when the mass of the fermionic singlet is less than the mass of the Higgs boson, the decay width does not increase even if the nonminimal coupling term between the Higgs field and gravity exists. The main AD channel is $\chi\to \nu_l h$ for fermionic singlet dark matter masses between 125~GeV and 700~GeV. The main AD channel is $\chi\to \nu_l hh$ for $700~\text{GeV}\leq m_\chi\leq 5.4~\text{TeV}$. Meanwhile, for fermionic singlet with heavier masses ($5.4~\text{TeV}\leq m_\chi\leq 10^3~\text{TeV}$), $\chi\to \nu_l hhh$ dominates the AD rate.

	Useful expressions of decay rates were also derived. As for scalar singlet dark matter, this paper derived an exact expression of the two-body decay width and an approximated expression of the three-body decay width. For fermionic singlet dark matter, this paper derived an exact expression of the two-body and three-body decay widths and an approximated expression of the four-body decay width. These expressions revealed that nonminimal coupling between the Higgs field and gravity could enlarge the decay width of gravitational dark matter. Comparison between MD width and AD width shows that the gravity-Higgs portal tends to dominate dark matter decay with a mass of several hundred GeV.

Many existing observations constrain the lifetime of dark matter to an extremely long duration. Therefore, this study conservatively required the dark matter lifetime to be greater than $10^{24}~\text{s}$ to obtain rough limits on nonminimal coupling constants. The results show that when we consider TeV physics, there is still a vast parameter space alive for both scalar and fermionic TeV dark matter candidates. Besides, dark matter particles with heavier masses are more constrained.


\bibliography{sunms}

\end{document}